\documentclass{ws-ijmpb}
\usepackage{amssymb}
\begin{document}

\markboth{I. Bena,  M. Droz, A. Lipowski}
{YANG-LEE FORMALISM FOR PHASE TRANSITIONS}

%
\catchline{}{}{}{}{}
%

\title{STATISTICAL MECHANICS OF EQUILIBRIUM AND
NONEQUILIBRIUM PHASE TRANSITIONS: \\ THE YANG-LEE FORMALISM}

\author{IOANA BENA}

\address{Theoretical Physics Department, University of Geneva,\\
 24, Quai Ernest Ansermet, CH-1211 Geneva 4, Switzerland\\
Ioana.Bena@physics.unige.ch}

%

\author{MICHEL DROZ}

\address{Theoretical Physics Department, University of Geneva,\\
 24, Quai Ernest Ansermet, CH-1211 Geneva 4, Switzerland\\
Michel.Droz@physics.unige.ch}

\author{ADAM LIPOWSKI}

\address{Quantum Physics Division, Faculty of Physics, A. Mickiewicz
University\\
Ul. Umultowska 85, 61-614 Poznan, Poland\\
lipowski@amu.edu.pl}

\maketitle

\begin{history}
\received{Day Month Year}
\revised{Day Month Year}
\end{history}

\begin{abstract}
Showing that the location of the zeros of the partition function can be used
to study phase transitions,  Yang and Lee initiated an ambitious and
very fruitful approach. We give an overview of the results obtained
using this approach. After an elementary introduction to the
Yang-Lee formalism, we summarize results concerning equilibrium
phase transitions. We also describe recent attempts and 
breakthroughs in extending this
theory to nonequilibrium phase transitions.
\end{abstract}

\keywords{Phase transitions; critical phenomena; Yang-Lee
formalism; equilibrium phase transitions; nonequilibrium phase
transitions.}

\section{Introduction}
\label{introduction}
Boiling of water, denaturation of a protein, or formation of a
percolation cluster in a random graph are examples of phase
transitions. Description and understanding of these ubiquitous
phenomena, that appear in physics, biology, or chemistry,
remains one of the major challenges of thermodynamics and statistical
mechanics.

The simplest and relatively well-understood phase transitions appear in
systems in thermal equilibrium with their environment. 
A prototypical example of such an {\em equilibrium phase transition} is the
liquid-gas transition. Crossing the coexistence curve, by varying
e.g., the temperature, one can move between vapor and liquid phases.
Such a transition is called first-order (or discontinuous) because
in the framework of thermodynamics this phenomenon is understood
in terms of thermodynamic potentials showing discontinuities in
their first derivatives with respect to the control parameters.
The coexistence line terminates at a particular point, the
critical point, where both phases become undistinguishable. Close
to this point, second derivatives of the thermodynamic potential
exhibit power-law singularities  and this indicates a second-order
(or continuous) phase transition. In the late fifties, giving a
microscopic description of first- and second-order phase
transitions became a challenge, that culminated in the development
of scaling theories and then renormalization group methods.
Although these techniques are computationally very effective and
reliable, they do not explain all the aspects of phase transitions. In
particular,  it is not clear how free energy can develop
singularities at the first-order phase transitions. To address
this problem, Yang and Lee developed an approach where the partition
function (that is used to calculate the  thermodynamic potentials) is
considered as a function of complex control parameters.
Singularities of the thermodynamic potentials, given as zeros of the
partition function, were then shown to accumulate exactly at the
transition point. This approach, that was later on generalized and extended to
various systems, provides a lot of information about equilibrium
phase transitions.

However, an equilibrium system is rather often only an abstract
idealization. Typically, a given system is not in equilibrium with
its environment, but is exchanging matter and/or energy with it;
fluxes are present in the system. The usual way to describe the
physics of such a system is to write down a master equation for
the time-dependent probability that a given mesoscopic state is
realized at time $t$. The physics is then embedded in the
transition rates between these mesoscopic states. In the long-time
limit the system eventually settles into a steady-state and,
depending on the values of some control parameters or the initial
configuration, different nonequilibrium steady states (NESS) can be
reached, and sometimes a breaking of ergodicity  takes place.
Upon varying some control parameters the system may change its
steady state and this is called a {\em nonequilibrium phase transition}.
In the particular situation in which the so-called detailed
balance condition is fulfilled, one recovers the case of
equilibrium phase transitions. Our understanding of NESS from a
microscopic point of view is not as advanced as for equilibrium.
Accordingly, it is legitimate to try to extend the Yang-Lee theory
to NESS. It was argued recently, and illustrated on several
examples, that Yang-Lee ideas should apply, at least to some
extent, to nonequilibrium phase transitions.

This review provides an overview of the Yang-Lee theory for
equilibrium phase transition and describe  recent advances in
applying it to nonequilibrium systems. In addition, we included an
extensive bibliographical list that offers an interested reader
the relevant references where one may find further details on a
more specific subject. Our presentation is elementary and
intuitive, and we omitted a number of technical details.

The structure of our paper is as follows. The first part of this
review is devoted to the description of the Yang-Lee theory for
equilibrium phase transitions. A first paragraph is devoted to the
question of the role played by the thermodynamic limit. Then the
general Yang-Lee theory for the grand-canonical partition function
is discussed. The celebrated Circle Theorem is revisited, as well
as its generalization to different models. Several aspects of the
characterization of the critical behavior in terms of the Yang-Lee
zeros are analyzed (density of zeros near criticality, Yang-Lee
edge singularity, finite size effects).
 A brief discussion of the relation between the Pirogov-Sinai
theory and the Yang-Lee zeros for first-order transitions is
given.  Then the problem of the Fisher zeros of the canonical
partition function is revisited. Finally, the Potts model is
discussed  in order to compare several possible types of
description of the phase transitions within the Yang-Lee
formalism.

The second part of this review is devoted to the question of
extending Yang-Lee formalism to NESS phase transitions. In a first
paragraph, the description in terms of a master equation is given,
and the question of detailed balance discussed. Then the problem
of a possible candidate for nonequilibrium steady-state partition
function is approached. Applications to driven-diffusive systems,
reaction-diffusion systems, directed percolation models, and
systems exhibiting self-organized criticality are reviewed. The
connection with equilibrium system with long-range interactions is
also considered. Finally, some conclusions and perspectives are
given in the last paragraph.

\section*{\bf{PART I: EQUILIBRIUM PHASE TRANSITIONS\vspace{0.5cm}\\}}
\label{equilibriumtransitions}

\section{Phase transitions and the thermodynamic limit}
\label{thermodynamiclimit}

When one varies a control parameter (for example, the temperature
or an external field), a physical system can exhibit a qualitative
change of its equilibrium state, i.e., it can undergo a {\em phase
transition}. In the framework of thermodynamics such a phase
transition shows up as a discontinuity or a singularity of
physical observables (e.g., specific heat, susceptibility, etc.)
as  functions of the control parameter. A characteristic
thermodynamic potential or some of its derivatives are
discontinuous or singular, thus {\em nonanalytic} at the
transition point.

How is it possible to understand this phenomenon from a
microscopic perspective? From the point of view of the equilibrium
statistical mechanics, the thermodynamic potentials can be
expressed in terms of some properties of the microstates in the
phase space of the system, that are determined by the microscopic
interactions between the constituents of the system (particles,
spins, etc...), as well as by the externally imposed constraints
-- like a fixed value of a control parameter. More explicitely,
the characteristic potential is proportional to the {\em logarithm
of the partition function} of the system. This partition function
is a sum of the statistical weights over different configurations
that are accessible to the system in the phase space under the
given constraints,  and these statistical weights are positively
defined analytic functions of the control parameter.~\footnote{For
example,  consider a system in equilibrium with a heat bath at
temperature $T$ (that represents a control parameter in this
case). The free energy (which is the characteristic potential)  is
proportional to the logarithm of the canonical partition function,
that is  defined as a sum over all the microstates $\alpha$ of the
Boltzmann weight $\exp(-E_\alpha/k_BT)$, where $E_\alpha$ is the
energy of the microstate $\alpha$.} If the sum contains a finite
number of such statistical weights, then the partition function is
finite, and its logarithm, i.e.,  the corresponding characteristic
potential is an analytic function of the control parameter, and
therefore no phase transition is possible. To obtain a
nonanalyticity of the characteristic potential,  the partition
function should thus necessarily become zero for a certain value
of the control parameter. However, as justified above, this cannot
happen in a finite system. A nonanalytic behaviour could only be
obtained for a system containing an infinite number $N$ of
constituents, and thus one has to work in the so-called {\em
thermodynamic limit}, in which both $N \rightarrow  \infty$ and
the volume of the system $V \rightarrow \infty$, and such that the
density $n=N/V$ is kept constant.

One may object that all the systems in nature are finite. However,
because of the huge number of its components (of the order of
Avogadro's number), any macroscopic system behaves,  from an
experimental point of view, like an infinite system  (except
eventually  in a very narrow domain in the vicinity of the
transition point). The effects of the finite-size of the system on
the appearance and properties of the phase transition  will be
subject of further discussion in Sec.~\ref{finitesize}.

To summarize, the presence of a phase transition, from a
statistical mechanics point of view, should be related to the
vanishing of the partition function for a certain value of the
control parameter.
 Thus one has to look for the zeros of the partition function, and
such zeros should only show up in the thermodynamic limit.

Let us briefly comment now on the properties of the interaction
potential between the constituents of the system that allow for
the existence of a correct thermodynamic limit.~\footnote{For
such a system, under some supplementary constraints on the
regularity of its frontier, taking the thermodynamic limit $N
\rightarrow \infty$,  $V \rightarrow \infty$  at fixed $n=N/V$
implies that: (i) The various statistical ensembles are
equivalent, and the relevant thermodynamic parameters are thus
uniquely defined. For example, the pressure has the same
thermodynamic limit, positively definite, in both the canonical
and grand-canonical ensembles. (ii)  The values of the
thermodynamic potentials (energy, free energy, grand-canonical
potential, enthalpy, etc.) {\em per particle} are finite. The
system enjoys the additivity property, which means that its total
Hamiltonian is the sum of the Hamiltonians of any combination of
macroscopic parts taken separately (i.e., one can safely neglect
the surface interaction energy between macroscopic component
parts) -- and so do all the thermodynamic potentials (free energy,
enthalpy, etc.). (iii) The thermodynamic stability of the system
is ensured, i.e., the thermodynamic potentials enjoy the required
conditions of convexity/concavity. For example, the
compressibility coefficient is positive, which means the pressure
is a non-decreasing function of the particle-number density $n$.
(iv) The effects of the boundaries (e.g., surface energy or
tension) are extinguished, and we speak only of bulk effects.}
Roughly speaking (see
Refs.~\refcite{vanhove49,ruelle63,fisher64,fisher65,ruelle69} for
further technical details),  for systems with two-body central
interactions, the interaction potential $u(r)$ (with $r$ the
distance between the particles)
has to obey the following three conditions:\\
(i)  The intermolecular forces have to approach zero rapidly
enough for large interparticle separations, i.e.,  in terms of
the interaction potential, $\left|u(r)\right|\leqslant
C_1/r^{d+\varepsilon}$ as $r\rightarrow \infty$ (with $C_1 >0$ and
$\varepsilon$ positive constants), where $d$ is the dimension of
the system.
Such systems are currently called {\em short-range interaction} systems.\\
(ii) $u(r)$ has to have a repulsive part increasing sufficiently
rapid at small interparticle distances (preventing the system
from collapse at high particle number densities). Throughout
this paper, we shall consider systems with a {\em hard-core}
interparticle repulsion at small distances, $u(r) = \infty$ for $r
< r_0 \sim b^{1/d}$ (where  $b$ is the single-particle excluded
volume), which means that a system of volume $V$ can accommodate a
{\em finite} maximum number of particles $M=V/b$.\footnote{As
shown in Ref.~\refcite{fisher64}, one can relax the hard-core
condition to $u(r)\geqslant C_2/r^{d+\varepsilon}$ as
$r\rightarrow 0$ (with $C_2$ and $\varepsilon$ positive
constants),
but  we will not consider this situation here.} \\
(iii) Finally, the interaction potential has to be everywhere bounded from below,
$u(r)\geqslant -u_0$  whatever $r$ (with $u_0$ a positive constant).

Note  a rather general result concerning the {\em one-dimensional
equilibrium systems with short range interactions}, namely van
Hove's theorem~\cite{vanhove49,ruelle69,landau80}, according to
which {\em no phase transition is possible} in such systems, in
contrast to what happens generically in  corresponding
nonequilibrium systems, see Ref.~\refcite{evans00} for a brief
review. A recent critical discussion~\cite{cuesta04} of this
result allowed to highlight some exceptions to this theorem, but
we shall not be concerned here with these rather pathological
situations.

Many physical situations can be modeled by interaction potentials
with the characteristics (i) -- (iii) above. There are, however, a
few exceptions. The most salient example is that of the systems
with long-range interactions (or non-additive systems),
which include, e.g., systems with gravitational or Coulombian
forces (see Ref.~\refcite{dauxois02} for a modern review of their
(thermo)dynamical properties studies). Such systems exhibit an
unequivalence of the statistical ensembles, and it is not clear
yet how to apply the concepts of the Yang-Lee theory (as described
below) to their phase transitions. We will not address them
further here. A limiting case of such systems is that of
mean-field  interactions, on which we will briefly comment  in
Secs.~\ref{circlegener}, \ref{fisher}, \ref{pottspotts}.

\section{Yang-Lee  zeros of the grand-canonical partition function: the general framework}
\label{generalframe}

We shall address here two problems, namely:\\
\\
{\bf (A) The location of the phase transition point}: How, when
knowing the partition function (in the thermodynamic limit), can
one locate a phase transition point
by investigating the zeros of this partition function with respect to the control parameter of the system.\\
\\
{\bf (B) The characteristics of the transition}: How can one
extract information on the nature of the phase transition (e.g.,
if it is a discontinuous or a continuous one)
from the properties and distribution of these zeros.\\

The presentation in this  section follows, with slight modifications, the
same general lines of thought as  in Ref.~\refcite{blythe03}. \\

{\bf (A) The location of the phase transition point.}
Let us start with this first problem and concentrate on the case of a
generic system with  short-range interactions as described above,
for which the thermodynamic limit is well-defined.
For concreteness, and following
the original papers of Yang and Lee~\cite{yang52,lee52},
we consider here the {\em grand-canonical} description
of the system. The corresponding partition function,
for a given volume $V$
and a fixed temperature $T$, is expressed as:
\begin{equation}
\Xi_V(T,z)=\sum_{m=0}^{M}Z_m(T){z^m}\,,
\end{equation}
where $M=V/b$ (with $b$ the single-particle excluded volume)  is the maximum
number of particles that can be accommodated in the system
(as determined by the hard-core part of the binary interaction potential);
$Z_m(T)$ is the canonical partition function of the system with
fixed number of particles $m$;
and finally $z$ is the {\em fugacity} of the system,
\begin{equation}
z=\exp(\mu/k_BT)\,,
\end{equation}
which is obviously a real, positive quantity expressed in terms of
$\mu$, the chemical potential of the system
in contact with an external reservoir of particles; $k_B$ is Boltzmann's constant.

One notices  that $\Xi_V(T,z)$ is a polynomial of $M$-th  order in the fugacity $z$.
Let us consider its {\em roots},
i.e., the solutions of  $\Xi_V(T,z)=0$;
according to the fundamental theorem of algebra, there are $M$ such roots,
$z_i=z_i(T)$, with  $i=1,...,M$.
Moreover, because the coefficients of all the powers of $z$ in
the expression of $\Xi_V(T,z)$ are real and positive, all these roots
$z_i(T)$
appear in {\em complex-conjugated pairs  in the
complex-fugacity plane, away from the real, positive semi-axis}.
One can express $\Xi_V(T,z)$ in terms of these roots,
\begin{equation}
\Xi_V(T,z)=\xi\,\prod_{i=1}^{M}\left[1-\frac{z}{z_i(T)}\right]
\end{equation}
(where $\xi$ is a multiplicative constant, that we shall
ignore in the foregoing), and the corresponding finite-size
grand-canonical potential
$\Omega_V=-k_B\,T \ln \Xi_V=-P_V\;V$ leads to the following
expression of the finite-size pressure $P_V$:
\begin{equation}
P_V(z)=k_B\,T\,\frac{1}{V}\sum_{i=1}^{M}\ln\left[1-\frac{z}{z_i(T)}\right]\,.
\label{pv}
\end{equation}

Note that throughout this Section we shall consider the {\em temperature $T$
as a fixed parameter}, and therefore all the quantities
like $P_V(z)$, $\rho(z)$,
$P(z)$, $\varphi(z)$,
$\psi(z)$, $\lambda(s)$, as well as ${\cal C}$ (see below) are {\em temperature-dependent}.

Let us  consider now the {\em complex extension of $P_V(z)$}, that
is defined through Eq.~(\ref{pv}) above for all {\em complex} $z$
{\em  with the exception of the points $z_i$}. From the standard
theory of the  complex-variable functions, one can deduce that
$P_V(z)$ is analytical (infinitely differentiable) over any region
of the complex-$z$ plane that is free of zeros of the partition
function. Therefore, a  nonanaliticity  of $P_V(z)$ at a complex
point $z_0$ appears if and only if in any arbitrarily small region
around   $z_0$ one finds at least one root of the partition
function (or, in other terms, if and only if there is an
accumulation of the roots of the partition function in the
vicinity of $z_0$). If, moreover, $z_0$ lies on the physically
accessible real positive semi-axis, this corresponds to a phase
transition in the system.

Once more, one realizes that such conditions cannot be
accomplished in a finite-size system. Let us  then turn to the
thermodynamic limit, where one  might eventually expect a possible
accumulation of the roots of the partition function towards the
real positive semi-axis.

As $V$ and therefore $M=V/b$ increase,
the location of the roots $z_i$  changes,
and in the {\em thermodynamic limit} they accumulate
in a certain region ${\cal C}$
of the complex-$z$ plane, with a local density $\rho(z)$.
Of course, in view of its significance,
$\rho(z)$ is a {\em real-valued, non-negatively defined function} on
${\cal C}$,  identically zero for any $z$ outside ${\cal C}$,
which is normalized as
\begin{equation}
\int_{\mathbf{ R}^2}\rho(z)\,dz=\int_{{\cal
C}}\rho(z)\,dz=\frac{M}{V}=\frac{1}{b}\,,
\label{normalization}
\end{equation}
and thus it is integrable over any bounded
region of the complex-$z$ plane.\footnote{As already mentioned in
Sec.~\ref{thermodynamiclimit}, we shall consider here only systems with a
hard-core interparticle repulsion at small distances, i.e.,
with a non-zero single-particle excluded volume $b$.
Most of the litterature on
the subject is limited to this situation;
see Ref.~\refcite{hauge63} for an exception.}
Moreover, the symmetry property of the roots with respect to the real axis
is preserved in the thermodynamic limit, and leads to the following property
of the density of zeros:
\begin{equation}
\rho(z)=\rho(z^*)
\label{symmetry}
\end{equation}
(where $(...)^*$ denotes complex-conjugation), i.e.,
${\cal C}$ is symmetric with respect to the real axis.
Note also that the region ${\cal C}$ depends on the
temperature $T$, and, of course,  its characteristics are specific  to each system.
For all the points $z$ outside the region ${\cal C}$,
one can define the thermodynamic limit
of the complex extension of the pressure,
$P(z)=\mbox{lim}_{V\rightarrow \infty}P_V(z)$, a {\em complex-valued} quantity
expressed as
\begin{equation}
P(z)=k_BT\int dz' \rho(z')\ln\left(1-\frac{z}{z'}\right)\,,
\label{pressure}
\end{equation}
which is, of course, a multi-valued function.
Let us consider its real part (up to a factor $k_BT$)
\begin{equation}
\varphi(z)\equiv\frac{1}{k_BT}\,{\cal R}eP(z)=\int dz' \rho(z')\ln\left|1-\frac{z}{z'}\right|\,.
\end{equation}
Since $\ln|z|$ (with $z=x+iy$; $x$, $y$ $\in \mathbf{R}$) is the
Green function of the two-dimensional Laplacian
$\Delta=\partial^2/\partial x^2+\partial^2/\partial y^2$, by
applying the Laplacian to the above equation, one finds that:
\begin{equation}
\rho(z)=\frac{1}{2\pi}\,\Delta \varphi(z)\,.
\label{potential}
\end{equation}
This means that $\varphi(z)$ is the electrostatic potential
associated to a distribution of charges of density $\rho(z)$.
This analogy is very useful, since it allows the  direct
transcription of well-known results from electrostatics. In
particular, since $\rho(z)$ is integrable even on regions
containing parts of ${\cal C}$, the function $\varphi(z)$ can be
extended through {\em continuity} over the entire complex-$z$
plane. The equipotential surfaces are thus given by
$\varphi(z)$=constant. Still in this analogy with electrostatics,
the imaginary part of $P(z)/k_BT$,
\begin{equation}
\psi(z)\equiv\frac{1}{k_BT}{\cal I}m P(z)
\end{equation}
(defined modulo $2\pi$) determines, through the condition
$\psi(z)$=constant, the lines of force of the electrostatic
field generated by the distribution of charges $\rho(z)$, and the
intensity of the field is given as $\mathbf{\bigtriangledown}
\varphi(z)$ and is, of course, discontinuous on ${\cal C}$, see
point (ii) below for further details. Note also that
\begin{equation}
\varphi(z)=\varphi(z^*) \quad \mbox{and} \quad \psi(z)=-\psi(z^*)\,,
\end{equation}
which result directly from the symmetry property (\ref{symmetry}) of the distribution of zeros.

Suppose now that the known partition function of the system
indicates us that  $\varphi(z)$ has, for example, two distinct
analytical expressions $\varphi_1(z)$ and $\varphi_2(z)$ in two
different regions of the complex-$z$ plane. But since $\varphi(z)$
has to be continuous over the entire plane, there should be a
matching region between these expressions, and, of course, this
region can be nothing else than ${\cal C}$. Indeed, since
$\varphi_{1,2}$ are different functions, there appears the
possibility of a nonanaliticity  of $\varphi$ (and thus of $P(z)$)
at the matching points. Then the location of ${\cal C}$ is given
by the condition:
\begin{equation}
\left.\frac{}{}\varphi_1(z)\right|_{{\cal C}}=\left.\frac{}{}\varphi_2(z)\right|_{{\cal C}}\;\;,\mbox{i.e.\,,}
\quad \left.\frac{}{}{\cal R}eP_1(z)\right|_{{\cal C}}=\left.\frac{}{}
{\cal R}eP_2(z)\right|_{{\cal C}}\,.
\label{realgen}
\end{equation}

If the region $C$ intersects the real positive semi-axis at some point $z_0$, and/or
there is an accumulation of zeros in the vicinity of $z_0$, then we have
a phase transition in the system,
and
\begin{equation}
{\cal R}eP_1(z_0)={\cal R}eP_2(z_0)\,,
\label{real}
\end{equation}
the real part of the complex-valued pressure is {\em continuous} at the transition point.

Of course, depending on the shape of ${\cal C}$, several
transition points may appear, for different values of the
fugacity; one can also encounter the phenomenon of multiple phase
coexistence -- like, for example, the appearance of triple points
through the coalscence, at some temperature, of two simple
(two-phase) transition points. Moreover, recall that ${\cal C}$
depends on the temperature, and from here there appears the
possibility of the existence of a critical  temperature $T_c$,
such that for any $T<T_c$ the zeros of the partition function
accumulate in the vicinity of the real positive semi-axis (i.e.,
there is a phase transition in the system), while for any $T>T_c$
the zeros of the partition function are no longer accumulating in
the vicinity of the real positive semi-axis, i.e., a phase
transition is no longer possible.

We thus answered the point (A) concerning the formal frame for the
description of the appearance
of the phase transitions --
at least, for the time being, in the grand-canonical ensemble.\\

{\bf (B) The characteristics of the transition}. In order to
answer the second question, namely the quantitative connection
between the nature of the phase transition and the distribution of
zeros of the partition function, let us suppose for the moment
that ${\cal C}$ is a {\em smooth curve} in the vicinity of the
transition point $z_0$. As it will be seen below, on some concrete
prototypical examples of physical systems, this is  rather often
the case for the zeros of the grand-canonical partition function
in the complex-fugacity plane -- the so-called {\em Yang-Lee
zeros}.

\begin{figure}[htb]
\centerline{\psfig{file=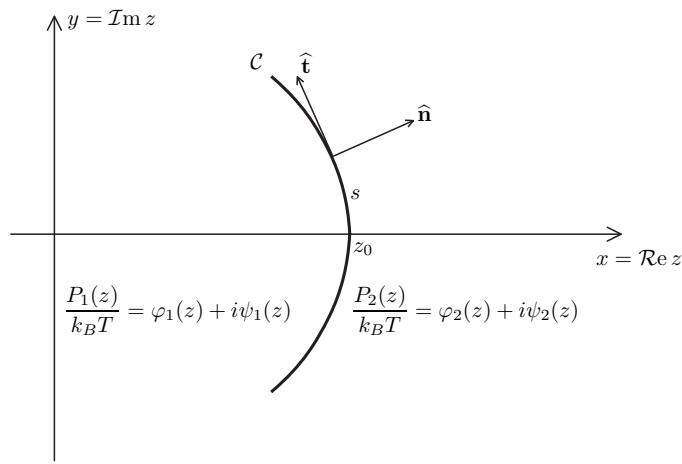,width=9cm}} \vspace*{8pt}
\caption{Schematic representation of the location ${\cal C}$ of
the Yang-Lee zeros in the vicinity of a transition point $z_0$,
meant to illustrate the notations in the main text.}
\label{figure1}
\end{figure}

Following the arguments in
Refs.~\refcite{blythe03,grossmann67,abe67I,abe67III,grossmann69I,grossmann69II,bestgen69},
let us consider a parametrization of the curve ${\cal C}$ in the
vicinity of $z_0$, with the parameter $s$ measuring the
anti-clockwise oriented distance from $z_0$ along the curve ($s=0$
at $z=z_0$), see Fig.~\ref{figure1}.

Let $\lambda(s)$ denote the line density of zeros along ${\cal C}$. Then, according to
Gauss' theorem, one has for a point $z$ on the curve
\begin{equation}
\left.\frac{1}{2\pi}\left[\boldmath{\nabla}\varphi_2(z)-\boldmath{\nabla}\varphi_1(z)\right]
\cdot \boldmath{\hat{n}}\right|_{{\cal C}}=\lambda(s)\,,
\end{equation}
where $\boldmath{\hat{n}}=\boldmath{\hat{n}}(s)$ is the unit vector normal to ${\cal C}$
(oriented from region ``1" to region ``2") at that point $z$. In view of the Cauchy-Riemann properties of the
analytical functions $P_{1,2}(z)$, one has
\begin{equation}
\left.\frac{}{}\boldmath{\nabla}\varphi_{1,2}(z)\cdot \boldmath{\hat{n}}\right|_{{\cal C}}=
\left.\frac{}{}\boldmath{\nabla}\psi_{1,2}(z)\cdot \boldmath{\hat{t}}\right|_{{\cal C}}
\end{equation}
where
\begin{equation}
\psi_{1,2}(z)=\frac{1}{k_BT}{\cal I}m P_{1,2}(z)
\end{equation}
are the imaginary parts of $P_{1,2}(z)/k_BT$ (defined modulo $2\pi$),
and $\boldmath{\hat{t}}=\boldmath{\hat{t}}(s)$ is the unit vector tangent to ${\cal C}$.
This leads finally to
\begin{eqnarray}
\lambda(s)=\left.\frac{1}{2\pi}\,\frac{d}{ds}\left[\psi_2(z)-\psi_1(z)\right]\right|_{{\cal C}}\;\;, {\mbox i.e.\,,}&&\;\; \lambda(s)=\left.\frac{1}{2\pi k_BT}\,\frac{d}{ds}\left[{\cal I}m P_{2}(z)-{\cal I}m P_{1}(z)\right]\right|_{{\cal C}}\nonumber\\
&&
\label{imaggen}
\end{eqnarray}
where $d/ds$ denotes the directional derivative along the curve. In particular, at the
transition point $z_0$, the discontinuity in the directional derivative of the
imaginary part of the complex-valued pressure is determined by the
local linear density of zeros,
\begin{equation}
\left.\frac{1}{2\pi k_BT}\,\frac{d}{ds}\left[{\cal I}m P_{2}(z)-{\cal I}m P_{1}(z)\right]
\right|_{z=z_0}=\lambda(0)\,.
\label{imag}
\end{equation}

The relationships~(\ref{realgen}), (\ref{real}), (\ref{imaggen}),
and (\ref{imag}) will allow us to establish the nature of the
phase transition, i.e., whether it is discontinuous (or
first-order) or continuous  -- i.e., second- or
higher-order in the classical {\em Ehrenfest classification
scheme}.

Let us consider the Taylor expansion around $z_0$
of the complex-valued pressure on both sides of the curve ${\cal C}$, i.e.,
\begin{equation}
\frac{1}{k_BT}P_{1,2}(z)=\frac{1}{k_BT}P(z_0)+a_{1,2}(z-z_0)+b_{1,2}(z-z_0)^2+{\cal O}((z-z_0)^3)\,,
\end{equation}
In order for the pressure to be real on the real $z$ axis, all the
coefficients of the development have to be real. From the
condition~(\ref{realgen}), one finds that the equation for the
curve ${\cal C}$ is given by
\begin{equation}
(a_2-a_1)(x-z_0)+(b_2-b_1)\left[(x-z_0)^2-y^2\right]+ {\cal R}e\left[{\cal O}(z-z_0)^3\right]=0\,,
\end{equation}
where $x$ and $y$ are the real, respectively the imaginary part of $z$. Several situations may appear.\\
(i) {\em First-order phase transition}. If $a_2 \neq a_1$  then
the complex-valued pressure has a discontinuity in its first
derivative and  the transition is  first-order. If, moreover,
$b_2\neq b_1$, then in the vicinity of the transition point $z_0$
the curve  ${\cal C}$ is a hyperbola,
\begin{equation}
y^2=(x-z_0)^2+\frac{a_2-a_1}{b_2-b_1}(x-z_0)
\end{equation}
whose tangent in $z_0$ is parallel to the imaginary axis, i.e.,
${\cal C}$  crosses the real axis smoothly at an angle $\pi/2$.
Using Eq.~(\ref{imag}) we find the density of zeros at $z_0$,
\begin{equation}
\lambda(0)=\frac{a_2-a_1}{2\pi}\,,
\end{equation}
i.e., the density of zeros at the transition point of a first-order phase transition
is  nonzero.\\
(ii)  {\em Second-order phase transition}. If $a_2=a_1$, but $b_2 \neq b_1$, then
the curve ${\cal C}$ obeys the equation
\begin{equation}
y=\pm(x-z_0)\,,
\end{equation}
i.e., in the vicinity of $z_0$ it consists of two straight lines that make an angle of
$\pm \pi/4$ with the real axis (and $\pi/2$ between them) and meet at $z_0$.
From Eq.~(\ref{imaggen}) we find that
\begin{equation}
\lambda(s)=\frac{b_2-b_1}{\pi}\,|s| + {\cal O}(s^2)\,,
\end{equation}
i.e., the desity of zeros is decreasing linearly to zero
when approaching  the transition point $z_0$ ($s=0$). \\
(iii) {\em Higher-order phase transitions}. If the discontinuities
appear at higher order in the derivatives of the complex-valued
pressure, then one can repeat the above type of reasoning to find
the equation of ${\cal C}$ and the density of zeros in the
vicinity of the transition point $z_0$. In general, if the
transition is of $n$-th order ($n \geqslant 3$), then the density
of zeros is zero at the transition point, $\lambda(s) \sim
|s|^{n-1}$, and the curve ${\cal C}$ does not cross smoothly
the real axis, but approaches it at an angle $\pm \pi/2n$ from
above and below.

Resuming, we managed therefore to respond the two main  questions:\\
(A) The accumulation of zeros of the partition function along the
 (physically acessible) real, positive semi-axis of the complexified fugacity $z$
 indicates the location of the phase transition point(s); \\
(B) The density of zeros near such an accumulation point
determines the order of the transition (according to Ehrenfest's
classification scheme)  at that point.

In this section we discussed the zeros Yang-Lee zeros of the
grand-canonical partition function in the complex-fugacity plane.
 One can, of course, address the following legitimate question:
how can one extend this type of formalism in order to study
phase transitions in other  statistical ensembles (e.g., the
canonical one). Indeed, in view of the equivalence of these
ensembles in the thermodynamic limit, one is entitled to expect
similar results on the location and characteristics of the phase
transitions whatever the ensemble used  in the description of the
system. Before addressing this important problem in
Sec.~\ref{fisher} below, let us add a few more relevant comments
on the Yang-Lee zeros.

\section{Grand-canonical partition function zeros and the equation of state}
\label{equationofstate}

We saw above that knowing the distribution $\rho(z)$
of the zeros of the grand-canonical partition function
in the thermodynamic limit, one can compute the
complex-valued pressure $P(z)$ through Eq.~(\ref{pressure}),
 \begin{equation}
 P(z)= k_BT \chi(z)\,,
 \label{part1}
 \end{equation}
with
\begin{equation}
\chi(z) \equiv \int dz'\,\rho(z')\,\mbox{ln}\left(1-\frac{z}{z'}\right)\,,
\label{part2}
\end{equation}
that is analytic everywhere outside ${\cal C}$.
 Also, the  complex-valued particle-number density
 $n(z)$ follows from the complex extension of
 \begin{equation}
 n(z)=\lim_{V \rightarrow \infty} \frac{\partial}{\partial \mbox{ln}(z)} \,
 \frac{1}{V}\,\mbox{ln}\,\Xi_V
  \end{equation}
 as:
 \begin{equation}
 n(z)= z\, \frac{d\chi(z)}{dz}=z\,\int dz'\,\frac{\rho(z')}{z-z'}\,,
 \label{part3}
 \end{equation}
for all complex $z$ outside ${\cal C}$. When $z$ is {\em real and
positive} (and outside the possible intersections of ${\cal C}$
with the real positive semi-axis that corresponds to
phase-transition points),  Eqs.~(\ref{part1}) and (\ref{part3})
(with $\chi(z)$ given by Eq.~(\ref{part2}) as a functional of
$\rho(z)$) are parametric expressions of the {\em equation of
state} of the system. Recall  also that $\rho(z)$ and ${\cal C}$,
and therefore  $\chi(z)$ are {\em temperature-dependent}, see
Sec.~\ref{generalframe}, and thus the structure of these
parametric equations is actually quite intricate. Of course, one
can, in principle, eliminate $z$ between the two equations and
find the explicit form of $P=P(n,T)$ (for the different regions of
the real-positive semi-axis of $z$ outside ${\cal C}$, i.e., free
of transition points).

However, one has to realize that the problem of determining the
distribution of complex  zeros $\rho(z)$ is an extremely difficult
task, even in the simplest known-cases like, e.g.,  a
one-dimensional lattice gas with nearest-neighbour attractive
interactions~\cite{lee52}, a gas of hard rods~\cite{hauge63}, a
mean-field lattice gas~\cite{katsura54,katsura55}. It is of
interest, therefore, to address  the following problem. Suppose
that we are given an certain equation of state $P=P(n,T)$; using
the parametrization of Eqs.~(\ref{part1}) and  (\ref{part3}), one
can find a closed nonlinear first-order differential equation for
the function $\chi(z)$, which, by integration, leads generically
to a functional equation of the form
\begin{equation}
{\cal F}(\chi, \,z)=0\,,
\label{functional}
\end{equation}
(for real positive $z$). Let us extend it, by definition, to the
whole complex-$z$ plane. Would it then be possible to solve it
and, once $\chi(z)$ is found, to determine the region ${\cal C}$
of accumulation of zeros of the grand-canonical partition
function? Supposing  that ${\cal C}$ is a smooth curve, would it
be then possible to find, through Eq.~(\ref{part2}) or
Eq.~(\ref{imaggen}), the corresponding density of zeros $\rho(z)$
of the grand-canonical partition function of the system?

This is what is currently called {\em the inverse problem} (in
analogy with the inverse problem in electrostatics), and it has
been addressed in several papers on various specific systems, see
Refs.~\refcite{pathria91,kortman71} (two models of Ising
ferromagnets), \refcite{hauge63}  (a Tonks gas of hard rods and a
gas with a weak long-range repulsion), \refcite{penrose68} (Tonks
gas),
\refcite{hemmer64,nilsen67I,ikeda74I,ikeda74II,ikeda74III,ikeda75}
(the van der Waals gas), \refcite{hemmer66} (a lattice gas with a
hard-core repulsion extended over several lattice sites),
\refcite{byckling65,niemeijer70} (Takahashi lattice gas),
\refcite{vandenberg76} (point particles with logarithmic
interactions), and even in an experimental setup, see
Ref.~\refcite{binek98} (for a two-dimensional Ising ferromagnet).

The answer to this issue  is related to the considerations in the
previous section and, being rather technical, will not be given
here in detail; we refer the reader to the set of four papers by
Ikeda, Refs.~\refcite{ikeda77,ikeda79,ikeda82I,ikeda82II},  for a
rigorous approach of the problem (and also some illuminating
examples, like various systems with circular distributions of
zeros, see Sec.~\ref{circle} below; the ideal Fermi-Dirac gas;
and the ideal Bose-Einstein gas). Roughly, an equation of
type~(\ref{functional}) has as solutions one or several
complex-valued  functions, which may have one or several  Riemann
surfaces (i.e., are in general multi-valued functions), with
branching points corresponding to $dz/d\chi=0$. One has to try to
match these different Riemann sheets,  along the curve ${\cal
C}$, so that several mathematically and physically necessary
conditions are fulfilled by the resulting patch-function
$\chi(z)$:  (i) The branching points belong to ${\cal C}$; (ii)
The real part of $\chi(z)$ ($\varphi(z)$ in the notations of the
previous section) has to be  continuos in the whole $z$ plane;
(iii) $\chi(z)$ is real and continuous on the real positive
semi-axis and has the value given by Eq.~(\ref{part1}); (iv) The
integration constant result from the  infinite-dilution (ideal
gas) limit,  $\chi(z) \sim z$ for $z \rightarrow 0$; (v) The
quantity $\rho(z)$, as defined by Eq.~(\ref{imaggen}), has to be
real-valued and non-negative for all $z$ on ${\cal C}$. The
discontinuities in the imaginary part of $\chi(z)$ ($\psi(z)$ in
the notations of the previous section) appear thus across the
charged contour ${\cal C}$ originating from matching the
different Riemann sheets of the multi-valued solutions of
(\ref{functional}). In the particular case when the solution of
(\ref{imaggen}) is a single multi-valued function, the domain
${\cal C}$ of the zeros of the partition function reduces to the
branching points. One has to note, however, that for a general
functional relation~(\ref{functional}) there is no guarantee of
the  uniqueness of the domain ${\cal C}$, and also that rather
often the explicit construction of ${\cal C}$, and thus the
calculation of $\rho(z)$ are not possible, except for limiting
cases, e.g., in the vicinity of the branching points.

\section{The lattice gas and the Ising model in a magnetic field}
\label{latticeising}

Before proceeding in the next sections with a discussion of the
domain ${\cal C}$
of accumulation of zeros of the grand-canonical partition function,
let us recall
a well-known classical result of equilibrium statistical mechanics, namely
that the problem of a discrete lattice gas  (with nearest-neighbour
interactions)
is mathematically equivalent to the problem of an Ising model
(with nearest-neighbour interactions) in an uniform magnetic field,
see, e.g., Refs.~\refcite{lee52,fisher65} for a detailed description.
In particular, the expression of the {\em grand-canonical}
partition function
of the lattice gas is identical to the expression of the
{\em canonical} partition function
of the Ising model in magnetic field
(at the same fixed temperature $T$).
The search of the complex roots of the grand-canonical
partition function of the gas in the complex-fugacity
plane amounts to the search of the complex roots of
the canonical partition function of the Ising model
in the plane of the complex variable $z=\exp(-2 H/k_BT)$
(in the appropriate units for the field and the temperature),
that is related to a {\em complex magnetic field intensity $H$}.

Therefore, based on this one-to-one correspondence,
in the foregoing we will speak either of a discrete
lattice gas, or of an Ising model in a magnetic field.

\section{Yang-Lee zeros and the transfer matrix formalism}
\label{transfermatrix}

For one-dimensional Ising lattice systems with finite-range interactions, in a magnetic field $H$,  it is
customary to compute  the partition function  using the {\em transfer-matrix} technique. The
canonical partition function  for $N$ spins can be written as:
\begin{equation}
Z_N=\lambda_1^N+\lambda_2^N+...+\lambda_k^N\,,
\end{equation}
where $\lambda_i$, with $i=1,...,k$, are the $k$ eigenvalues of
the $k\times k$ transfer matrix of the system. Besides being
dependent on the reduced coupling constants (i.e., the coupling
constants divided by $k_BT$)  between the spins, these eigenvalues
depend on the value of the magnetic field $H$, or, equivalently,
on the  fugacity $z=\exp(-2 H/k_BT)$, $\lambda_i=\lambda_i(z)$,
$i=1,...,k$. Let us consider the free energy per spin, $F=-k_B
T\,\displaystyle{\mbox{ln}Z_N}/{N}$, and its extension to the
plane of the complex fugacity. In the thermodynamic limit, only
the eigenvalue of highest modulus contribute to the free energy.
Suppose now that in two different regions of the complex-$z$ plane
two different eigenvalues, $\lambda_1(z)$, respectively
$\lambda_2(z)$  assume the largest modulus. Since the real part of
the free energy (per spin) has to be continuous throughout the
whole $z$ plane -- see the discussion in Sec.~\ref{generalframe}
in the light of the correspondence between Ising systems and
lattice gases, Sec.~\ref{latticeising} -- it follows  that the
location ${\cal C}$ of the zeros  of the partition function in the
complex-$z$ plane is given by the condition of matching of the
modulus of these two eigenvalues of the transfer matrix:
\begin{equation}
\left.\frac{}{}\left|\lambda_1(z)\right|\right|_{{\cal C}}=\left.\frac{}{}\left|\lambda_2(z)\right|\right|_{{\cal C}}\,,
\end{equation}
which is the equivalent of Eq.~(\ref{realgen}) in the present
frame. Of course, in view of the van Hove theorem
(Sec.~\ref{thermodynamiclimit}) for the systems with short-range
interactions  that we are considering here, no phase transition
is possible, i.e., ${\cal C}$ does not have any accumulation point
on the real positive semi-axis for such systems. This matching
condition between eigenvalues might be useful for the construction
of ${\cal C}$, see, e.g., Ref.~\refcite{katsura72I,yamada81} for a
few examples. A more rigorous discussion of this type of approach
for a wide class of lattice gases can be found in
Ref.~\refcite{elvey74}.

\section{The Circle Theorem for the Yang-Lee zeros}
\label{circle}

The general frame discussed in Section~\ref{generalframe}
above is applicable
to all the systems  -- within, of course, the mentioned
restrictions that refer, essentially,
to the existence of the thermodynamic limit,
see~Sec.~\ref{thermodynamiclimit}. However, as already mentioned,
the shape of the domain ${\cal C}$ of the accumulation of zeros
of the grand-canonical partition function in the complex-fugacity plane,
besides being
{\em temperature-dependent}, is {\em specific to each system},
and a priori one might expect it to have a rather intricate structure.
It is therefore a rather amazing
result that, for a quite general class of systems, this region
${\cal C}$ can be proven rigorously to {\em lie on the unit circle}
$|z|=1$ in the complex-fugacity plane. This is the celebrated
Circle Theorem, that was first demonstated by Yang and Lee,
see Ref.~\refcite{lee52}, for a rather restricted class of
Ising systems. The beauty and simplicity of this result
attracted a lot of research afterwards, and led to its
generalization to many other situations and models,
see below. Note, however, that no general statement can be
made about the density of zeros on the unit circle, which
is characteristic to each system and, of course,
temperature-dependent.

\subsection{The original Yang-Lee Circle Theorem}

The system originally considered by Yang and Lee
(see Refs.~\refcite{lee52,griffiths72,lebowitz68})
consists of an ensemble of $N$ Ising $1/2$-spins $\sigma_i$ ($i=1,...,N$):\\
(i) that are placed on a $d$-dimensional lattice, \\
(ii) with pair ferromagnetic interactions between them,
of coupling constants
$J_{ij}\geqslant 0$ ($i\neq j=1,...,N$), \\
(iii) that are subject to a (eventually inhomogeneous) magnetic field.\\
The corresponding Hamiltonian\footnote{Of course, alternatively, one can consider the equivalent lattice gas model, which corresponds to an uniform magnetic field.}
\begin{equation}
{\cal H} = -\sum_{i <j} J_{ij} \sigma_i \sigma_j - \sum_i H_i\sigma_i
\end{equation}
leads to a canonical partition function $Z_N=Z_N(z_1, z_2, ..., z_N)$
that is a multinomial in the fugacities
\begin{equation}
z_i=\exp(-2H_i/k_BT)
\end{equation}
($H_i$ being the value of the magnetic field acting on the spin
$\sigma_i$). In the particular case of a homogeneous magnetic
field $H_i=H$ for all $i$, the partition function is simply a
polynomial of degree $N$ in the fugacity $z=\exp(-2H/k_BT)$.
Under the supplementary hypothesis\\
(iv) $|z_i| \leqslant 1$ for all $i=1,...,N$\\
one can prove the {\em Circle Theorem}: the zeros of the partition
function in the $2N$-dimensional space of complex $z_i$  lie all
on the unit circle,
\begin{equation}
Z_N=0 \rightarrow |z_1|=|z_2|=...=|z_N|=1\,,
\end{equation}
which amounts at saying that these roots correspond to strictly
imaginary (or zero) values of the magnetic field intensity $H_i$,
$i=1,...,N$. In the particular case of a homogeneous magnetic
field, all the $N$ roots of $Z_N=0$ lie on the unit circle $|z|=1$
in the complex-z plane, i.e., appear for $N$ strictly imaginary
(or zero) values of $H$. The theorem remains valid in the
thermodynamic limit, when the zeros form a dense set ${\cal C}$ on
the unit circle, characterized by the linear density $\lambda$. Of
course, an immediate consequence of this theorem is the well-known
classical result that a phase transition might appear in such a
system, at a fixed finite temperature $T$, only in zero magnetic
field.

Note that:\\
(i) The dimensionality of the lattice does not play any role in
the demonstration of the theorem, which is thus valid in any
dimension. Moreover, the size and
the structure of the lattice, its regularity, periodicity, translational
symmetry do not play any role either in the obtention of the result.\\
(ii) The ferromagnetic interactions are not restricted to first
neighbours. However, they should decrease sufficiently rapidly
with the distance in order to ensure the existence of the
appropriate thermodynamic limit, see the discussion in
Sec.~\ref{thermodynamiclimit}.

Indeed, the demonstration of this theorem~\cite{lee52}, that will
not be reproduced here, rely on some properties of the Hamiltonian
that are independent of the dimension and of the range of the
interactions; besides the ferromagnetic character of the interspin
interaction, an essential necessary ingredient is the {\em
spin-reversal symmetry} of the system, i.e., the invariance of its
Hamiltonian with respect to global inversion of the spins and of
the (inhomogeneous) magnetic field, which leads to the following
symmetry property of the canonic partition function:
\begin{equation}
Z_N(z_1,z_2,...,z_N)=z_1z_2...z_N Z_N(z_1^{-1},z_2^{-1},...,z_N^{-1})\,.
\end{equation}
Later on Asano, in Refs.~\refcite{asano70I,asano70II}, introduced a new, more general  technique for the study of the location of the zeros of the grand-canonical partition function. This method was often used to prove the Circle Theorem for various systems (see below), but it was also extended by Ruelle, see Ref.~\refcite{ruelle71,ruelle73},  so as to permit statements about regions other than the unit circle.

\subsection{Generalizations of the Circle Theorem to other systems and models}
\label{circlegener}

We shall briefly present below some of the systems for which
it was shown, using various methods, that the Circle Theorem is valid.\\
\\
{\bf (A) Modified Ising ferromagnets}, which include:\\
(i) {\em Ising ferromagnets of arbitrary spin}, see
Refs.~\refcite{asano68,suzuki68I,griffiths69I}. The basic idea for
the extension is that the higher-spin ``atom" is a cluster of
$1/2$-spin ``atoms", that are coupled together through a suitable
ferromagnetic interaction. This representation leads to an
effective, temperature-dependent, $1/2$-spin Ising Hamiltonian and
from here to the Circle Theorem.
Once demonstrated this result for any finite value of the spins, one can extend it
to the classical limit or infinite-spin limit, see Ref.~\refcite{harris70}.\\
(ii) {\em Diluted Ising ferromagnets}.  Consider a lattice whose
sites are either occupied by a spin $1/2$ (with a probability
$p$), or by a non-magnetic atom (with probability $1-p$); $p$
is thus equal to the concentration of spins on the (infinite)
lattice. As shown in Refs.~\refcite{suzuki68II,kunz70}, for a
ferromagnetic coupling between the spins, and in a homogenous
magnetic field $H$,
the zeros of the partition function correspond to purely imaginary
(or zero) values of $H$. This result can be extended to higher-spin cases.\\
(iii) {\em Ising models with many-spin interactions}. Ising
lattice systems with interactions involving three or more sites
-- i.e., whose  Hamiltonian comprises interaction terms  of the
form $-J_{ij...k}\sigma_i \sigma_j ...\sigma_k$, where $i\neq j
\neq ... \neq k$ correspond to different lattice sites, and
$J_{ij...k}$ is the coupling constant between these sites  --
have been used to model a variety of physical situations,  like,
for example, some binary alloys~\cite{styer86}, lipid
bilayers~\cite{scott88}, and gauge-field theory
models~\cite{kogut75}. These systems have rich phase diagrams, and
Monte-Carlo results indicate the presence of phase transitions at
nonzero values of the magnetic field for a number of these
systems, see, e.g., Refs.~\refcite{chin87,heringa89}. However, in
view of their complexity, very few analytically rigorous results
are known. In Ref.~\refcite{suzuki71} it is shown that for a set
of many-spin interactions of finite range, under some conditions
on the coupling constants, and also some spin-inversion symmetry
conditions, the Circle Theorem holds (at least)  up to a certain
temperature (that is determined by the coupling constant and does
not depend on the number of spins in the system). The example of
four-spin interactions shows explicitly that the zeros leave the
unit circle at large-enough temperatures. Qualitative arguments
suggest that this is the case
for other systems with multi-spin interactions, see also Ref.~\refcite{monroe91}.\\
(iv) {\em Ising models on hierarchical lattices}. In
Ref.~\refcite{southern87} a rescaling formalism is used in order
to study the location of the Yang-Lee zeros for $1/2$-spin and
$1$-spin Ising chains on two- and three-dimensional Sierpinski
gaskets, as well as on 3-simplex and 4-simplex lattices. The
Circle Theorem is shown to hold for nearest-neighbour
ferromagnetic interactions, but is no longer valid at high
temperatures when four-spin interactions are included, see also
point (iii) above. References~\refcite{bosco87,monroe95} address
the problem of the Yang-Lee zeros of the Ising nearest-neighbour
ferromagnetic model on a Cayley tree. The Julia set of the
renormalization transformation of the model gives the
thermodynamic limit of the partition function zeros distribution.
This one lies on the unit circle, and  is multi-fractal in nature,
with a temperature-dependent generalized dimension; below $T_c$
the Julia set is the unit circle (i.e., has dimension $D=1$), and
above $T_c$ the Julia set is a Cantor set (with a dimension
$D(T)<1$ decreasing with the temperature) on the unit circle. When
multi-site interactions
are included, these zeros leave the unit circle, see Ref.~\refcite{ananikian00}.\\
(v) {\em Aperiodic Ising models}.
References~\refcite{baake95,simon95,grimm97} discuss various one-
and two-dimensional examples of Ising spins with nearest-neighbour
interactions on aperiodic lattices. The interaction constant
can take only two values, $J_{a,b}$, distributed on the lattice
according to some generating rule. In $1D$ case, for example,
the actual distributions of the two coupling constants $J_{a,b}$
along the chain is determined by an infinite ``word" in the
letters ``a" and "b" which is obtained as the unique limit of
certain two-letter substitution rules (like, e.g., the Fibonacci
or the Thue-Morse rules). If both $J_{a,b}$ are positive
(ferromagnetic coupling), then, although the zeros still lie on
the unit circle, their distribution has a fractal structure (that,
of course, disappears when $J_a=J_b$), as shown by the integrated
density of the zeros along the circle. Thus such systems may be
regarded as
intermediate between regular and hierachical models.\\
(vi) {\em Mean-field Ising model}. In
Refs.~\refcite{katsura54,katsura55,saito61,abe67II}, is addressed
the problem of the Yang-Lee zeros for the Temperley-Husimi  model
of a lattice gas, or for the equivalent mean-field Ising model in
a magnetic field, which, of course, is known to exhibit a
first-order phase transition, with a critical temperature $T_c$.
The partition function  zeros lie on the unit circle in the
complex-fugacity plane, and one can compute analytically the
corresponding density. A direct comparison with Mayer's cluster
expansion theory~\cite{mayer37,mayer40,mayer42} shows that the
non-analiticity point in Mayer's series does not coincide with the
true transition point as obtained from the Yang-Lee theory, see
also related comments in, e.g.,
Refs.~\refcite{yang52,pathria91,toda98}. See also
Ref.~\refcite{gulbahce04} for an approach using the Yang-Lee zeros
of a restricted partition function for the study of the metastable
states of long-range
(and their mean-field limit) Ising models. \\
\\
{\bf (B) Heisenberg spin models}. The  technique of studying the
Yang-Lee zeros  that was introduced by Asano in
Refs.~\refcite{asano70I,asano70II}  was also used by Suzuki and
Fisher in Ref.~\refcite{suzuki71}, their studies being devoted to
the  extension of the Circle Theorem to quantum $1/2$-spin systems
in a magnetic field. More precisely, they considered the following
{\em fully anisotropic} Heisenberg Hamilonian for a system of $N$
spins of components $\sigma_i^{x,y,z}$, $i=1,...,N$,  in a
magnetic field:
\begin{eqnarray}
{\cal H}=&-&\sum_{i<j} \left(J_{ij}^x \sigma_i^x \sigma_j^x+J_{ij}^y \sigma_i^y \sigma_j^y+J_{ij}^z \sigma_i^z \sigma_j^z\right) \nonumber\\
&-&\sum_i \left(H_i^x \sigma_i^x+H_i^y\sigma_i^y+H_i^z\sigma_i^z\right)\,,
\label{anisotropic}
\end{eqnarray}
where $J_{ij}^{x,y,z}$ are the (anisotropic) coupling constants
and $H_{i}^{x,y,z}$ are the (non-uniform) components of the
magnetic field in the $x,y$ and $z$ directions. Consider that
$H_i^z \equiv H\mu_i$, with  $\mu_i$  non-negative constants (for
all $i=1,...,N$), and that the coupling constants obey the
following set of inequalities
\begin{equation}
J_{ij}^z \geqslant |J_{ij}^x| \quad \mbox{and} \quad J_{ij}^z \geqslant |J_{ij}^y| \quad
\mbox{for all}\;\; i\neq j=1,...,N\,,
\end{equation}
which correspond to the so-called~\cite{suzuki71} {\em
ferromagnetic Heisenberg model with dominant $z-z$ coupling}. Then
one can prove that the zeros of the partition function
$Z_N=\mbox{Tr}\left[\exp(-{\cal H}/K_BT)\right]$ as a function of
$H$ lie all on the imaginary-$H$ axis for all $N$, and do so also
in the thermodynamic limit. This result was generalized further in
Ref.~\refcite{suzuki71} for quantum models with spin greater than
$1/2$ (note that some preliminary results have already been
obtained in Ref.~\refcite{suzuki69}), and also to the classical
Heisenberg system of infinite spins, see
Ref.~\refcite{harris70}; see also Ref.~\refcite{kunz70} for a
temptative relaxation of the corresponding conditions
(\ref{anisotropic}) on the coupling constants in the classical
limit. In Ref.~\refcite{harris70} it was also shown that the
Circle Theorem holds for a {\em two-sublattice antiferromagnet of
arbitrary spin, in terms of the
staggered magnetic field intensity.}\\
\\
{\bf (C) Monomer-dimer models}. Consider a lattice of $N$ sites or
vertices, connected through $N(N-1)/2$ edges; one associates to
the edge connecting the vertices $i$ and $j$ a weight $w_{ij}$,
and to a vertex $i$ a weight $x_i$ (of course, $w_{ij}$ and $x_i$
are non-negative numbers for all $i \neq j = 1,...,N$). A
dimer-monomer covering of this lattice corresponds to (i) placing
dimers on the edges such that no two dimers share a common vertex
(i.e., no vertex has more than one dimer); (ii) placing monomers
on all the remaining vertices, i.e., the vertices  that are not
adjacent to a dimer. The statistical weight of such a covering is
then the product of all the weights $w_{ij}$ of the edges covered
by the dimers times the weights $x_i$ of all the vertices covered
by a monomer. The partition function
$Z_N=Z_N(\left\{w_{ij}\right\}, \left\{x_i\right\})$ of the system
is then the sum of the weights of all the possible coverings.
These models correspond to a huge variety of physical situations,
see, e.g., Ref.~\refcite{heilmann72} for an overview.

References~\refcite{heilmann70,heilmann71,heilmann72,kunz70,gruber71}
address the problem of the zeros of the partition function of the
monomer-dimer system. It is shown that, for given non-negative
$w_{ij}$, $Z_N$ cannot be zero if ${\cal R}e(x_i) >0$ for all
$i=1,...,N$ or  if ${\cal R}e(x_i) <0$ for all $i=1,...,N$. In
particular, if $x_i \equiv x$ for all $i=1,...,N$,  this leads to
the conclusion that the zeros of $Z_N$ appear on the
imaginary-$x$ axis (or at $x=0$), a result that holds also in the appropriate
thermodynamic limit.  This corresponds to the absence of a phase transition
in the monomer-dimer system, except, eventually, at $x=0$, which corresponds to
a maximum density of dimers. Connections with Ising and Heisenberg spin models are also discussed.\\
\\
{\bf (D) Ferroelectric models}. In
Refs.~\refcite{suzuki71,katsura70,chang71} it was shown that the
Slater-type models for ferroelectricity~\cite{lieb72I} can be
mapped onto Ising models with at most four-spin interactions.
Therefore, the Yang-Lee zeros lie on the unit circle at low
temperatures, but leave it at the critical temperature $T_c$. A
modified Slater model, and an antiferromagnetic model are also
discussed from the perspective of the Yang-Lee theory, and their
distributions of zeros on the unit circle, for
finite-size lattices, are investigated numerically.\\
\\
{\bf (E) Quantum fields}.
References~\refcite{simon73,dunlop75,newman74,newman75,newman76,lieb81}
show that some Euclidian quantum fields can be approximated by
generalized  Ising models on a lattice, with a given
probability distribution of the spins (that may be discrete or
continuos, but has to obey certain symmetry properties), and with
a self-interaction of the spins. In particular, the Yang-Lee
Circle Theorem is shown to hold for the considered systems,
see the above-cited references for further details.\\

For all these systems for which the Circle Theorem holds, using
the notations in Sec.~\ref{generalframe}  one can say that the
curve ${\cal C}$ lies on the unit circle $|z|=1$, and can be
parametrized by the trigonometric angle $\theta$. Correspondingly,
the  density of zeros  $\lambda=\lambda(\theta)=\lambda(-\theta)$
(in virtue of the symmetry property~(\ref{symmetry})), and it is
nonzero on two arches of the unit circle defined by  $\theta_{0}
\leqslant | \theta| \leqslant \pi$, where $\theta_{0}>0$ is a
function of temperature (see also Sec.~\ref{edge}). The
normalization condition~(\ref{normalization}) for the density of
zeros reads simply
\begin{equation}
2\int_{\theta_{0}}^{\pi} \lambda(\theta) \,d\theta =b\,.
\label{norm1}
\end{equation}

The complex pressure~(\ref{pressure}) becomes
\begin{equation}
P=k_BT \int_{\theta_{0}}^{\pi} \lambda(\theta)\,\mbox{ln}(z^2-2z\cos \theta+1)\,d\theta\,,
\end{equation}
and the complex particle-number density~(\ref{part3}) is
\begin{equation}
n=2z\,\int_{\theta_{0}}^{\pi}  \lambda(\theta)\,\frac{z-\cos \theta}{z^2-2z\cos \theta+1}\,d\theta\,.
\end{equation}
Note, however, that the density of zeros $\lambda(\theta)$ is
known analytically in very few cases (e.g., the one-dimensional
Ising model with nearest-neighbour interactions, see
Ref.~\refcite{lee52}; the mean-field Ising model,
Refs.~\refcite{katsura54,katsura55}), and was computed numerically
in few other cases (see Ref.~\refcite{kortman71} for such a
calculation for the Ising ferromagnets on a two-dimensional square
lattice, and on a three-dimensional diamond lattice,
respectively), and thus the practical use of these relationships
is rather restricted.

\subsection{Griffiths inequalities}

Note also a very important by-product of the Circle Theorem,
namely the demonstration of various {\em Griffiths-type
inequalities}, i.e., inequalities between spin-correlation
functions, see
Refs.~\refcite{griffiths69I,asano70I,asano70II,asano70III,simon73,dunlop75,newman75,newman76,dunlop79,sokal81}.
The simplest to demonstrate are the Griffiths-Kelly-Sherman
\cite{griffiths67I,griffiths67II,griffiths67III,kelly68} (GKS)
inequalities. They were first obtained using the Circle Theorem in
Refs.~\refcite{griffiths69I,asano70III}, for an Ising $1/2$-spin
ferromagnet in a magnetic field $H \geqslant 0$, as:
\begin{equation}
\langle \sigma_A \sigma_B\rangle \geqslant 0 \quad \mbox{(the first-type GKS inequality)}
\end{equation}
\begin{equation}
\langle \sigma_A \sigma_B \sigma_C \sigma_D\rangle \geqslant  \langle \sigma_A \sigma_B\rangle \,\langle \sigma_C \sigma_D \rangle
\quad \mbox{ (the second-type GKS inequality)}\,,
\end{equation}
where $\sigma_{A,B,C,D}$ are the products of the spins  inside the
(multi-site) regions $A,B,C,D$ of the lattice, respectively;
$\langle ... \rangle$ denotes the mean over the statistical
ensemble. These inequalities were  generalized, on the basis of
Yang-Lee Circle Theorem, to higher-spin
systems\cite{griffiths69I}, Heisenberg
ferromagnets\cite{asano70I,asano70II}, and systems with arbitrary
even spin distributions\cite{dunlop75}.

The Griffiths-Hurt-Sherman\cite{griffiths70} (GHS) inequality
\begin{equation}
\langle \sigma_A \sigma_B \sigma_C \rangle \leqslant \langle \sigma_A \rangle
\langle \sigma_B \sigma_C \rangle + \langle \sigma_C \rangle
\langle \sigma_A \sigma_B \rangle + \langle \sigma_B \rangle
\langle \sigma_C \sigma_A \rangle  -2 \langle \sigma_A \rangle
\langle \sigma_B \rangle \langle \sigma_C \rangle
\end{equation}
was demonstrated for various systems, using the Circle Theorem, in
Refs.~\refcite{simon73,newman75,newman76}. It has several
interesting implications, the concavity of the average
magnetization as a function of $H$; the monotonicity of the
correlation length as a function of the magnetic field; the
well-known result on the absence of phase-transition for $H >0$;
some inequalities for the critical exponents; see, e.g.,
Refs.~\refcite{ellis76,sokal81}.
 More complicated inequalities on the correlation functions,
involving the Circle Theorem, are treated in
Refs.~\refcite{newman75,dunlop79}.

\section{Further results on the Yang-Lee zeros location}
\label{other}

As already mentioned, the papers by Asano,
Refs.~\refcite{asano70I,asano70II}, and later on their elegant
generalization by Ruelle, Refs.~\refcite{ruelle71,ruelle73}, lead
to a  criterion for the location of the Yang-Lee zeros in the
complex-fugacity plane for a fairly large number of systems. This
criterion, as well as ad-hoc methods adapted for the system under
study (e.g., the transfer-matrix formalism for the one-dimensional
lattice systems, see Sec.~\ref{transfermatrix}), and sometimes
numerical results allowed to approach several systems and to draw
some general conclusions. A first remark would be that usually the
distribution of Yang-Lee zeros in the complex-fugacity plane is
well-behaved, i.e., in the thermodynamic limit
the zeros are distributed densely on an ensemble of smooth curves.\\
\\
{\bf (A)}  {\bf Ising systems with multi-spin interactions}. A  rather
general result refers to the fact that the Circle Theorem does not
hold  at high temperatures, see points (iii) and (iv)  for
modified Ising models in the previous section.
A  rigorous study of the regions in the complex-magnetic plane
that are {\em free}
of zeros of the partition function  is carried
in Ref.~\refcite{monroe91} for two very general classes of
multisite interaction
systems, namely (i) systems having ferromagnetic interactions
involving even numbers of sites, and (ii) systems with, again,
ferromagnetic interactions involving even numbers of sites
and, in addition, either ferromagnetic or antiferromagnetic
interactions involving odd numbers of sites. Each site may
interact with a finite number of other sites.
As for the case of the Circle Theorem, the dimension and the
specific type of lattice do not come into play. It is shown that:
(i) For systems of the first type, there is  an interval of the
real $H$ axis, $\left(-C(T), C(T)\right)$ (with $C(T)>0$ that depends on the
strength of the ferromagnetic couplings)
outside  which there is no phase transition. $C(T)$ goes to
zero as $T\rightarrow 0$, and hence the interval shrinks to $H=0$.
(ii) For systems of the second type, when the interaction
involving odd numbers of sites is
antiferromagnetic, there is an interval on the
real $H$  axis $(-\infty, -C(T))$ in which no phase transition occurs;
when the interaction involving odd numbers of sites is ferromagnetic,
there is no phase transition for real $H \in (C(T), \infty)$.
Here again $C(T)$ is a positive parameter, depending on
temperature and on the
interaction constants, that goes to zero
in the zero-temperature limit. These general results put clear
constraints on the region of the  plane temperature-magnetic
field where one is entitled to look for a possible phase transition
for such systems with mul;ti-spin interactions.
\\
\\
{\bf (B) Ising systems with antiferromagnetic interactions} or,
equivalently, {\bf lattice gases with repulsive interactions}.
Various Ising lattice systems with antiferromagnetic interactions
have been considered in Refs.~\refcite{hauge63} (a Tonks gas of
hard rods and a gas with a weak long-range repulsion),
\refcite{penrose68} (Tonks gas),
 \refcite{hemmer66} (a lattice gas with a hard-core repulsion
extended over several lattice sites), \refcite{nilsen67II} (a
hard-core lattice gas on an $M \times \infty$ square lattice),
\refcite{lebowitz68,suzuki70} (Ising antiferromagnet),
\refcite{harris70} (a two-sublattice antiferromagnet),
\refcite{lieb72II} (Ising ferromagnet on a bipartite lattice),
\refcite{katsura71,katsura72I} (Ising lattices with various
combinations of ferromagnetic and antiferromagnetic couplings),
\refcite{katsura72II} (Fisher and Temperley lattice gases),
\refcite{katsura72III} (one-dimensional $XY$ model),
\refcite{ohminami72} (antiferromagnetic Husimi-Temperley model),
\refcite{heilmann71} (lattice gases with all the zeros of the
partition function on the real negative semi-axis of the
fugacity), \refcite{runnels72} (various lattice systems of rigid
molecules), \refcite{matveev96I} (various Ising antiferromagnets),
\refcite{kim04I} (numerical results on finite square-lattice Ising
antiferromagnets). Different approaches  are used, that lie from
the very abstract rigorous mathematical ones, to approximate
analytical ones in solving the inverse problem
(Sec.~\ref{equationofstate}), and to numerical methods. The
generic emerging result is that, due to the presence of the
antiferromagnetic coupling (or, equivalently, of the repulsive
part of the lattice gas interparticle interaction), a part of the
Yang-Lee zeros lie on the real negative semi-axis of the fugacity
$z$ (for finite systems, as well as in the thermodynamic limit,
when eventually these zeros occupy densely a whole portion of the
negative semi-axis). The other details of the distribution of the
zeros are, of course, specific to the details of the considered
model (i.e., range of interactions, whether the coupling is mixing
ferromagnetic and antiferromagnetic features, etc.).
The locus of zeros and  even the density of zeros were sometimes computed analytically.\\
\\
{\bf (C) Degenerated Ising spins.} In Ref.~\refcite{yamada81} it
is discussed the case of a a one-dimensional, nearest-neighbour
interactions, spin-$1$ Ising system that has a {\em spin
degeneracy}, i.e., for which each spin variable $S$ can take
either of the values values $S=1, 0, -1$ with a certain weight
(degeneracy) $g(1), \,g(0), \,g(-1)$. It is shown that the
corresponding Yang-Lee zeros do not lie, in general, on the unit
circle in the complex-fugacity plane. The conditions under which
one recovers the Circle Theorem are discussed,
see also Refs.~\refcite{griffiths69I,suzuki73,yamashita78}.\\
\\
{\bf (D) Van der Waals gas}.  Several papers were devoted to this
prototypical model of non-ideal gas, that can be seen as the
continuum limit of a lattice gas with weak, long-ranged attractive
forces,  see Ref.~\refcite{hemmer66}.
References~\refcite{hemmer64,nilsen67I,ikeda74I,ikeda74II,ikeda74III,ikeda75}
addressed the problem of the distribution of Yang-Lee zeros,
showing that: (i) For infinite temperature the zero distribution
is located on part of the real negative semi-axis of the fugacity;
(ii) With decreasing temperature the distribution branches off the
real axis, circumventing the origin symmetrically, on both sides;
(iii) Below the critical temperature $T_c$ the distribution forms
a closed curve around the origin, with a diameter decreasing
exponentially to zero as $T \rightarrow 0$; (iv) An additional
tail of the distribution  remains on the negative real axis, but
with a density of zeros going linearly to zero as $T \rightarrow
0$. This tail is connected to the repulsive-core singularity of
the van der Waals interaction potential, see Ref.~\refcite{park99}
for a discussion. The density of zeros was computed analytically
in some limiting cases, e.g., in the vicinity
of the branching points. \\
\\
{\bf (E)  $\pm J$ Ising model for spin glasses}.
Reference~\refcite{ozeki88} discusses a numerical technique for
obtaining the distribution of Yang-Lee zeros of the disordered
symmetric $\pm J$ Ising model (i.e., an Ising  lattice for which
the coupling constant between two neigbour sites is chosen at
random as $J$ or $-J$), in two and three dimensions. This
distribution of zeros, which determines the analytical properties
of the configuration-averaged free energy, is the superposition of
the zeros of the partition function corresponding to each
configuration of the coupling constants. The zeros are not
distributed on smooth curves, and it seems that in the
thermodynamic limit they occupy (densely or not, the question is
not clarified yet) a whole region of the complex-$z$ plane. The
computation of the corresponding configuration-averaged free
energy allows to approach  some problems of this short-range spin
glass model -- like the existence of the Griffiths singularity (a
non-analytic behaviour in the paramagnetic phase of a random or
diluted Ising ferromagnet, see Ref.~\refcite{griffiths69II}) and
of  the  Almeida-Thouless transition line,
Ref.~\refcite{almeida78}.


\section{Yang-Lee zeros and the critical behaviour}
\label{critical}

For simplicity, the considerations in this section address the
particular case of a system for which the Circle Theorem  is
valid. However, they can be adapted, with some modifications,
 to more general locations of Yang-Lee zeros. For example, instead of the Yang-Lee edge angle $\theta_0$ in
 Sec.~\ref{edge}, one should consider, in general, the smallest distance between the ensemble ${\cal C}$
 of zeros and the  critical point.

\subsection{Density of zeros near criticality}
\label{critdens}

In the vicinity of a critical point the thermodynamic quantities,
as well as the correlation functions exhibit power-law behaviours
associated with a set of critical exponents. For example, for an
Ising model one has:
\begin{itemize}
\item For the (zero field) specific heat
\begin{equation}
c(t) \sim t^{-\alpha}, \quad {\rm for~} t>0
\end{equation}
and
\begin{equation}
c(t) \sim {\vert t \vert}^{-\alpha'}, \quad {\rm for~} t<0\,;
\end{equation}
\item For the spontaneous magnetization
\begin{equation}
m(t) \sim {\vert t \vert}^{\beta}, \quad {\rm for~} t<0 \,;
\end{equation}
\item For the isothermal susceptibility in zero field
\begin{equation}
\chi(t) \sim t^{-\gamma}, \quad {\rm for~} t>0
\end{equation}
and
\begin{equation}
\chi(t) \sim {\vert t \vert}^{-\gamma'}, \quad {\rm for~} t<0 \,;
\end{equation}
\item For the field dependence of the magnetization at criticality
\begin{equation}
m(h,t=0) \sim h^\delta\,;
\end{equation}
\item For the correlation length
\begin{equation}
\xi(t) \sim t^{-\nu}, \quad {\rm for~} t>0
\end{equation}
and
\begin{equation}
\xi(t) \sim {\vert t \vert}^{-\nu'}, \quad {\rm for~} t<0\,;
\end{equation}
\item Finaly, for the long-distance spin-spin correlation function at criticality :
\begin{equation}
G_2(r,t) \sim \frac{1}{r^{(d+2-\eta)}}\,,
\end{equation}
\end{itemize}
where $t=(T/T_c -1)$ is the reduced temperature ($T_c$ being the
critical temperature). The nine critical exponents $\alpha,
\,\alpha',\, \beta, \,\gamma, \,\gamma',\, \delta,\, \nu,\, \nu'$,
and $\eta$ are not independent, and the scaling laws reduce the
number of independent exponents to two.\cite{stanley87}

In the framework of the Yang-Lee theory, all the above quantities
can be expressed in terms of the density of zeros, e.g.,
$\lambda(\theta, t)$ for an Ising-like system for which the Circle
Theorem is valid. In the critical region, $\lambda(\theta, t)$
should take a particular scaling form in order to reproduce the
above behaviours. A large body of work has been devoted to the
study of this scaling form, see
Refs.~\refcite{suzuki68I,abe67IV,abe67V,suzuki67I,suzuki67II,suzuki68III,bessis76},
and the main result is that:
\begin{equation}
\lambda(\theta, t)= |t|^\beta \psi(\theta t^{-(\beta+\gamma)})\,,
\end{equation}
in the region of the unit circle $\theta_{0}(t) \leqslant |\theta|
\leqslant \pi$ where the density of zeros is different from zero
(see Eq.~(\ref{norm1}) and above it); $\psi(x)$ is a
positively-defined function, that is finite and nonzero at
$x=0^{\pm}$. Thus the critical exponent $\beta$ can be extracted
from the scaling behavior of $\lambda(\theta, t)$ in the vicinity
of the critical point.

\subsection{Yang-Lee Edge Singularity}
\label{edge}

Let us consider again a model for which the Circle Theorem is
valid, as for example an Ising-like model. For a temperature
$t>0$, a gap will show up in the density of zeros $\lambda(\theta,
t)$. No zeros are present in an interval
$[-\theta_0(t),\,\theta_0(t)]$ or, correspondingly, in a purely
imaginary external field interval $[-iH_0, +iH_0]$ (with
$H_0=k_BT\theta_0/2$ in the appropriate units, see
Sec.~\ref{latticeising}). This gap reduces to zero when the
temperature descends to the critical value and below. Kortman and
Griffiths~\cite{kortman71} were the first to point out the
interest to investigate the behavior of $\lambda(\theta, t>0)$ for
$\theta$ close to the points $\pm \theta_0(t)$, called the {\em
Yang-Lee edges}. A large literature addressed this question, see
Refs.~
\refcite{park99,baker79,wang98,fisher80,kurtze78,kurtze79,glumac91},
leading to the following picture. The edge singularity should be
regarded merely as an usual critical point occuring with a purely
imaginary field $iH$. It turns out that
\begin{equation}
\lambda(\theta, t>0) \sim {\vert \theta-\theta_0(t)  \vert}^\sigma, ~{\rm for~}
\theta  \to \theta_0(t)^+\,.
\end{equation}
Renormalization group arguments demonstrate that the appropriate
model describing the edge singularity is the so-called $\phi^3$
field theory.\cite{fisher78} The exponent $\sigma$ is simply
related to the thermodynamic exponent $\delta$:
\begin{equation}
\sigma=\delta^{-1}
\end{equation}
The upper critical dimension (above which mean-field predictions
are correct) is $d_u=6$. Moreover, the above power law behavior is
valid for all temperatures $T > T_c$, and also for all ${\cal
O}(n)$ models. Indeed, the presence of the magnetic field breaks
the original ${\cal O}(n)$ symmetry and thus the behavior becomes
Ising-like. Note moreover that the critical theory of the Yang-Lee
edge in two dimensions corresponds to a rather simple realization
of conformal symmetry.\cite{cardy85}

Therefore the behaviour of the density of zeros near the critical
edge provides a second independent critical exponent $\delta$. The
knowledge of $\beta$ and $\delta$, and of the scaling laws
characterizes completely the critical behaviour of the system.

\section{Finite-size scaling  and Yang-Lee zeros}
\label{finitesize}

As we have seen above, the density of zeros near the critical edge
and at criticality contains all the information concerning the
critical behavior of a given system. However, an analytic
expression for the density of zeros can be obtained only for some
particular simple systems. For more complicated cases, the density
of zeros has to be computed numerically on finite systems. It is
thus important to understand the role played by the finiteness of
the systems.

The finite-size scaling theory provides a powerful tool in
interpreting finite-size systems data. The theory was first
developped for continuous phase transitions based on
phenomenological arguments.\cite{fisher71} A more modern
presentation is based on a renormalisation group
approach.\cite{barber83} The key hypothesis is based on the
premise that only two different scales matter, namely: $\xi$, the
correlation lenght in an infinite system, and $L$, the
characteristic linear extent of the finite-size system. The case
of finite-size effects in first-order phase transition is somehow
more subtle as discussed by Fisher and Berker,
Ref.~\refcite{fisher82}. However, scaling at a first-order
transition can be treated in a similar manner as the one for a
second-order transition with the temperature and magnetic
anomalous dimensions, $y_t$ and $y_h$, assuming the maximal values
$y_t=y_h=d$, where $d$ is the dimension of the system.

For a finite-size system of size $L$ the singular part of the magnetization has the form
\begin{equation}
m(t,h,L)=L^{-d+y_h}\, m(tL^{y_t}, hL^{y_h})\,
\end{equation}
where
$h\equiv 2H/k_BT$. Then the density of zeros scales as
\begin{equation}
\lambda(\theta, t, L)=L^{-d+y_h} \,\lambda(\theta L^{y_h}, tL^{y_t})\,.
\end{equation}
At the critical temperature $t=0$ one has
\begin{equation}
\lambda_c(\theta,L)=L^{-d+y_h}\lambda_c(\theta L^{y_h})\,,
\end{equation}
which implies for the infinite system
\begin{equation}
\lambda_c(\theta) \sim |\theta|^{1/\delta}\,.
\end{equation}

For $t> 0$, the angle of the Yang-Lee edge (see the section above)
for finite systems, $\theta_0(t,L)$, scales as
\begin{equation}
\theta_0(t,L)=L^{-y_h}\theta_0(tL^{y_t})\,,
\end{equation}
leading,  for $L\rightarrow \infty$, to
\begin{equation}
\theta_0 \sim t^{\beta+\gamma}\,.
\end{equation}

Finite-size scaling for the density of Yang-Lee zeros
$\lambda(\theta,t,L)$ has been discussed by Alves  {\em et
al}\cite{alves00I}  for the three-dimensional Ising model, by
Kenna and Lang~\cite{kenna91,janke02II,kenna93} for the
four-dimensional Ising model and for the ${\cal O}(N)\, \phi^4$
theory at the upper critical dimension~\cite{kenna04}. Creswick
and Kim~\cite{creswick97,creswick95} have analyzed the critical
properties of the two-dimensional Ising model by computing exactly
the partition function of systems of sizes $4\leqslant L\leqslant
10$ and by extrapolating the data using a Bulirsch-Stoer (BST)
algorithm.\cite{bulirsch91} Janke and
Kenna~\cite{janke01I,janke04} have recently developed a novel
numerical technique which allows to determine both the latent heat
in the case of a first-order transition and the specific heat
exponent $\alpha$ in the case of a second-order transition. The
main point is the study the so-called {\em cumulative distribution
of zeros}, that is defined as:
\begin{equation}
\Lambda_L(r,t)=\int_0^r\lambda(s,t,L)ds\,.
\end{equation}
This approach has been successfully applied to the study of $2d$
and $3d$ Ising models, $d=2, \,q=10$ Potts model (see
Sec.~\ref{potts}),  and lattice gauge theories.


\section{Pirogov-Sinai
theory and Yang-Lee zeros for first-order phase transitions}
\label{pirogov-sinai}

A recent series of papers, see
Refs.~\refcite{lee94,lee96,biskup00,biskup04I,biskup04II},
addresses the problem of  a more direct connection of the
Yang-Lee zeros with the (complex extension of the) free energy and
the discontinuities of its derivatives, for {\em
general}~\footnote{``General" means here systems with no
particular symmetry properties like, e.g., the spin-inversion
symmetry invoked in deriving the Circle Theorem} lattice spin
models in a magnetic field, that present first-order phase
transitions. The departure point is the Pirogov-Sinai formalism
for the phase diagram of the system, see
Refs.~\refcite{pirogov75,pirogov76}. Its basic hypothesis are the
existence of a finite number of ground states (which are the
thermodynamic phases) of the system, and of the availabilty of an
appropriate contour representation, i.e., the fact that the
partition function can be written as the sum over the partial
partition functions of each phase. Roughly speaking, this
corresponds to the fact that, for a system with  ${\cal P}$
possible phases (${\cal P}=1,2,3,...$), the probability
distribution for the {\em order parameter} (the magnetization)
conjugated to the external field $H$ has ${\cal P}$ peaks; the
partition functions corresponding to the different phases can be
evaluated from the position and width of these peaks (which depend
on the control parameters temperature and magnetic field $H$). The
total partition function is then expressed as
\begin{equation}
Z_N(z)=\sum_{l =1}^{\cal P} g_l\exp[-Vf_{l}(z)/k_BT] \,+{\cal
O}(\exp(-L/L_0))\,,
\label{pirogov}
\end{equation}
where $N$ is the number of spins; $f_{l}(z)$
is the complex extension of the thermodynamic free energy
per unit volume of phase $l$,
as a function of the complex fugacity $z=\exp(-2H/k_BT)$;
$g_l$ is the degeneracy of phase $l$;
$V=L^d$ is the volume of the lattice of linear extension
$L$ and dimension $d$;
and $L_0$ is of the order of the correlation length.
Let us suppose here, for simplicity, the absence of
triple or higher-order
coexistence points (the theory can be extended to include
these cases, too), i.e., for any values of the control parameters
$(T, H)$ the
probability distribution function of the order parameter
has no more than two
peaks.
Then by inspection of Eq.~(\ref{pirogov}) one realizes that the
zeros of the partition function in the complex-$z$ plane
arise, up to order ${\cal O}(\exp(-L/L_0))$,
from a destructive interference of pair of terms of the sum,
$q_l\exp(-Vf_l/k_BT)$ and $q_p\exp(-Vf_p/k_BT)$,
$l \neq p=1,...,{\cal P}$.
So, each zero of $Z_N(z)$ in the complex-fugacity plane lies
in a vicinity of the order ${\cal O}(\exp(-L/L_0))$
of a solution of the equations
corresponding to this condition of destructive interference:
\begin{eqnarray}
&&{\cal R}e f_{l}(z)-(k_BT/V)\mbox{ln}g_l=
{\cal R}e f_{p}(z)-(k_BT/V)\mbox{ln}g_p\,,\nonumber\\
&&V/k_BT [{\cal I}mf_l(z)-{\cal I}mf_p(z)]=\pi \,\mbox{mod}\,2\pi\,,
\quad l \neq p=1,...,{\cal P}\,.
\end{eqnarray}
In the thermodynamic limit, the zeros concentrate asymptotically on the
{\em phase coexistence curves} ${\cal R}e f_{l}(z)=
{\cal R}e f_{p}(z)$,
and the corresponding  local density of zeros is given by
$\rho(z)=({2\pi k_BT})^{-1}\,{d}[f_l(z)-f_k(z)]/dz$.
Practically, this means that the zeros of the partition function can
indeed be expressed in
terms of the complex free energy and of the discontinuities in its
derivatives.
References~\refcite{lee94,biskup00} illustrate the application of this
rather abstract-looking method to three examples, the low temperature
Ising and Blume-Capel, and the $q$-state Potts model in the limit of large $q$.

\section{Fisher zeros of the  canonical partition function}
\label{fisher}

In a review article of 1965, Ref.~\refcite{fisher65}, Fisher
suggested the analysis of the phase trasitions in the frame of the
{\em canonical} ensemble, through the study of the distribution of
the zeros of the canonical partition function in the plane of the
complex temperature -- i.e., what we call today {\em Fisher's
zeros}. Indeed, in view of the equivalence of the various
statistical ensembles in the thermodynamic limit, one should be
able to equally characterize a phase transition in any of these
ensembles. As in the grand-canonical case, the Fisher  zeros will
lie off the real positive temperature semi-axis for a finite
system, but may close up on this semi-axis in the thermodynamic
limit; the point $T_c$ where a limiting line of zeros cuts the
semi-axis will locate a {\em critical point}. Fisher illustrated
this idea in Ref.~\refcite{fisher65} on the two-dimensional
square-lattice Ising model, with isotropic nearest-neighbour
interactions, in the absence of the external magnetic field, whose
partition function is known analytically since Onsager, see
Ref.~\refcite{onsager44}. Fisher found  that the zeros of the
canonical partition function lie on an unit circles in the plane
of the complex variable $v=\mbox{sinh}(2J/k_BT)$, where $J>0$ is
the ferromagnetic coupling constant, and $T$ is the {\em complex}
temperature. These circles cut the real-$v$ axis in $v=\pm 1$; the
point $v=+1$ corresponds to the ferromagnetic transition, and
$v=-1$  corresponds to the antiferromagnetic phase transition
point (in agreement with the $\pm J$ symmetry property of the
square lattice in zero magnetic field).

The concept of Fisher zeros (both in the absence and in the
presence of a magnetic field), which  seems to follow closely that
of the Yang-Lee zeros studies, was  rapidly set-up in its general
frame, including the relevance of the thermodynamic limit, the
location of the critical point,
Refs.~\refcite{fisher65,jones66,majumdar66}, and the
characterization of the transition, see
Refs.~\refcite{grossmann67,abe67I,abe67III,grossmann68,grossmann69I,grossmann69II,bestgen69}.
The (often tacit) underlying assumption of these abstract studies
on the characteristics of the transition is that in the
thermodynamic limit the Fisher zeros fall on smooth,
complex-conjugate curves,  at least in some vicinity of the
critical point. In particular, provided that this is indeed the
case,  Itzykson {\em et al}, see
 Ref.~\refcite{itzykson83}, proved two very strong general
statements for Ising systems, namely: (i) These complex-conjugate
curves in the absence of magnetic field, in the vicinity of the
critical point, form an angle with the real temperature axis that
is an universal known function of the critical exponent $\alpha$
(of the specific heat) and of  the ratio of the specific heat
amplitudes below and above the critical point. (ii) The angle
(with respect to the real temperature axis) at which these zeros
curves (in the vicinity of the critical point) depart when a
real, weak magnetic field is switched on is an universal function
of the critical exponents $\beta$ (of the spontaneous
magnetization) and $\delta$ (of the relation between field and
magnetization at the critical temperature).

However, despite some exceptions (like, e.g.,  the above-cited
Onsager-Ising model), the above  assumption on the location of
Fisher zeros is rarely fulfilled, see, e.g.,
Ref.~\refcite{vansaarloos84} for a discussion of this point. The
rule seems to be that the Fisher zeros in the thermodynamic limit
occupy densely some whole areas in the complex-temperature plane,
delimited by curves on which the density of zeros may diverge.
Nothing like a Circle Theorem or an  Asano-Ruelle type of argument
exists for Fisher zeros, because the canonical partition function
with respect to a variable of the type $v=\mbox{sinh}(2J/k_BT)$
(where $J$ is a typical interaction constant of the model, and $T$
is the complex temperature)  does not have the polynomial
structure the grand-canonical partition function has with respect
to the fugacity $z$.
In fact, there are three elements that render the study of Fisher zeros more complicated than that of Yang-Lee zeros:\\
(i) The first element is that the location of the Fisher zeros is
strongly dependent on the details of the system (dimension,
interaction between the components of the system, etc.). For
example, Fisher zeros for the Onsager-Ising model have completely
different location than the Fisher zeros of the  corresponding
anisotropic ferromagnet. This general feature is, of course, in
relation
with the fact that the critical temperature $T_c$ is a quantity that varies from one system to another.\\
(ii)  In general, the zeros for a {\em finite} system are not
necessarily located in the region of the complex plane occupied
densely by the Fisher zeros in the thermodynamic limit. See, for
example, the case of a mean-field Ising ferromagnet, in
Refs.~\refcite{caliri84,glasser86,glasser87,bena03,greenblatt04,gulbahce04},
where the location of the Fisher zeros in the thermodynamic limit
is interpolated from that corresponding to
larger and larger finite systems.\\
(iii) The third point is that rather always the location of the
Fisher zeros, even in the thermodynamic limit, is not
well-behaved, i.e., it does not correspond to smooth curves in
the complex-$v$ plane, but rather to a combination of curves and
densely-covered regions of the plane (like, for example, for the
very simple example of an anisotropic ferromagnet on a square
lattice, see, e.g., Ref.~\refcite{vansaarloos84}).

These elements explain the scarcity of exact analytical results on
the Fisher zeros (as compared to the Yang-Lee zeros), the
predominance of numerical results, and thus the importance of
finite-size scaling arguments in the study of the characteristics
of the transition by mean of Fisher zeros. Without entering into
details, let us give a few examples of the systems that were
studied
in the literature:\\
\\
{\bf (A) Ising models with nearest-neighbour ferromagnetic
interactions},
including: \\
(i) Two-dimensional finite $m\times n$ isotropic square lattices,
in zero magnetic field, Ref.~\refcite{katsura67,abe70}; see also
Ref.~\refcite{brascamp74} for a discussion of the role of boundary
conditions. Reference~\refcite{lu98} generalizes the result of
Fisher~\cite{fisher65} on the circular location of  the zeros to
$m \times \infty$
rectangular lattices with (asymmetric) self-dual boundary conditions, and closed form expression of the density of zeros are obtained for $m=1, 2$.\\
 (ii) The Onsager-Ising model in an external, symmetry-breaking
magnetic field of {\em complex} value $H/k_BT=\pm i\pi/2$, whose
exact partition function in the thermodynamic limit was first
presented by Lee and Yang in Ref.~\refcite{lee52}, was approached
from the point of view of the complex-temperature zeros in
Refs.~\refcite{matveev95,matveev96II}. The density of Fisher zeros
is discussed in Ref.~\refcite{lu01}
(also for the cases of triangular, honeycomb, and Kagom\'e lattices).\\
(iii) The corresponding problem in a {\em real} nonzero magnetic
field was approached numerically and analytically (with
low-temperature series expansions) in
Refs.~\refcite{suzuki70,matveev96II}. It was found that the
density of Fisher zeros diverges at a non-physical critical point,
the {\em Fisher edge singularity} that can be considered  as the
equivalent
of  the Yang-Lee edge singularity (see Sec.~\ref{critical}), as discussed also in Ref.~\refcite{kim02}.\\
(iv) The loci of the Fisher zeros in the thermodynamic limit of
two-dimensional systems on honeycomb, triangular, diced, and
Kagom\'e lattices, in the absence of the magnetic field, were
described in Ref.~\refcite{abe91}.
Reference~\refcite{lu01} discusses the corresponding density of the Fisher zeros.\\
(v) Two-dimensional Ising systems on square, triangular, and honeycomb  lattices are investigated both in the isotropic and {\em anisotropic} cases, for finite systems, and in the thermodynamic limit, in zero magnetic field, in Refs.~\refcite{vansaarloos84,stephenson84,stephenson86I,stephenson86II,stephenson87,stephenson88I,stephenson88II}. It is concluded that, in general, the Fisher zeros occupy densely the complex-$v$ plane in the thermodynamic limit, and the corresponding density is also discussed.\\
(vi) Fisher zeros for finite three-dimensional isotropic,
cubic-lattice Ising ferromagnets, in zero magnetic field,
as well as estimates  (based on finite-size scaling arguments) of their densities, were discussed, at various levels, in Refs.~\refcite{ono68,suzuki70,pearson82,alves00I}.\\
\\
{\bf (B) Higher-spin Ising model}. Ising models of spin S=1, 3/2,
2, 5/2 and 3 have been calculated  in Ref.~\refcite{kawabata69}
for a two-dimensional Ising model on a square lattice, for various
lattice sizes, in zero magnetic field. The temperature-dependence,
as well as the asymmetry in the specific heat amplitude above and
below the critical point
are estimted from the Fisher zeros distribution.\\
\\
{\bf (C) Mean-field spin models}. The problem of the Fisher zeros
for mean-field $1/2$-spin Ising models -- including the relation
with critical exponents, finite-size scaling aspects, the use of a
partial distribution function in order to characterize the
metastable behavior, the relationship with long-range interaction
systems -- was approached in
Refs.~\refcite{abe67II,caliri84,glasser86,glasser87,bena03,greenblatt04,gulbahce04}.
The exact locus of the Fisher zeros for the mean-field spherical model was obtained in Ref.~\refcite{kastner05}. See also Ref.~\refcite{alves00II} for an Ising-type mean-field model for a polypeptide with helix-coil transition.\\
\\
{\bf (D) Antiferromagnetic Ising models}.
References~\refcite{suzuki70,kim05} obtains numerically the
location of the  Fisher zeros for a finite Ising antiferromagnet
on a square lattice, for different values of the external magnetic
field and different extensions of the lattice. It is inferred that
in the thermodynamic limit
these zeros have an accumulation point on the real axis, corresponding to  the antiferromagnetic critical point.\\
\\
{\bf (E)  Heisenberg ferromagnets}.
References~\refcite{kawabata70I,kawabata70II,kawabata70III}
discuss the thermodynamic functions (such as energy, specific
heat, magnetization,
and susceptibility) that have been computed numerically, using Fisher zeros distributions,  for finite three-dimensional Heisenberg models with different anisotropy constant parameters, including both ferromagnetic and antiferromagnetic couplings.\\
\\
{\bf (F) Ising models on hierarchical lattices}. The location,
fractal nature,  and density of Fisher zeros for Ising models on
various such lattices are studied in
Refs.~\refcite{derrida83,lee99} (diamond hierarchical lattice), \refcite{debell84,ananikian00} (Cayley trees), \refcite{southern87,burioni99,liaw02} (Sierpinski gaskets and 3- and 4-simplex lattices). We also refer the reader to a short, comprehensive review of Itzykson  {\em et al.} on this subject, Ref.~\refcite{itzykson85}.\\
\\
{\bf (G) Ising models on aperiodic lattices}.
References~\refcite{baake95,simon95,grimm97,grimm02} discuss the
Fisher zeros for several such one- and two-dimensional Ising
aperiodic lattices, with
coupling constants generated according some predefined algorithms, see also Sec.~\ref{other}.\\
\\
{\bf (H) $\pm J$ Ising spin model}. In Refs.~\refcite{bhanot95,saul95} there are presented numerical results on the corresponding Fisher zeros for two- and three-dimensional lattices of different sizes. More extended simulations in Ref.~\refcite{damgaard95} seem to indicate a fractal nature of the distribution of zeros.\\
\\
{\bf (I) Random energy model}. The random energy model is one of
the simplest, exactly-solvable,
{\em disordered} model  that contains some physics of the spin glasses, see Ref.~\refcite{derrida81}. It is equivalent to a spin model with multispin interactions exhibiting a low-temperature spin glass phase. Its Fisher zeros were studied numerically and analytically in Refs.~\refcite{moukarzel91,derrida91}, and they were find to occupy densely lines and extended areas of the complex-temperature plane.\\

Most of the above-cited references invoke finite-size scaling
arguments in order to infer the properties of the Fisher zeros
(location zone, position of the critical point of the transition,
density, etc.) in the thermodynamic limit from those obtained
numerically for finite-size systems. A long list of references
concentrate precisely on the {\em finite-size scaling} theory for
Fisher zeros  (and, sometimes, they address also the same problem
for the Yang-Lee zeros), see
Refs.~\refcite{bhanot87,bhanot90,kenna91,kenna93,creswick95,janke01I,janke01II,janke02I,janke02II,janke02III,janke02IV,janke04,kenna04,huang04}.

\section{Potts model: what to complexify~?}
\label{potts}

The Potts model  (so named after R. B. Potts who first studied it
in its 1951 Ph.D. thesis at Oxford) is a generalization of the
Ising model to more-than-two-component spins, and, due to its
complex behavior, as well as the various experimental
realizations, has been a subject of intense research during the
last thirty years, see Ref.~\refcite{wu82} for a review. The
$q$-state Potts model consists of $q$-components spins, i.e.,
spins that can point in the $q$ symmetric directions of a
hyperthetraedron  in $(q-1)$ dimensions, described by the unit
vectors $\bf{s}^{\alpha}$, $\alpha=0,...,q-1$; these spins are
placed on a lattice, and the corresponding Hamiltonian ${\cal H}$
of the system in the presence of a magnetic field $H$  is
expressed as:
\begin{equation}
{\cal H}=-\varepsilon \sum_{\langle ij \rangle } \delta({\bf s}_i,\,{\bf s}_j)
- H \sum_{i} \delta({\bf s}_i,\,{\bf s}^{Q})\,.
\end{equation}
Here $\bf{s}_i$ is the direction of the $i$-th spin as discussed
above; $\varepsilon$ is the coupling constant (which is {\em
ferromagnetic} if $\varepsilon >0$, and {\em antiferromagnetic} if
$\varepsilon <0$); $\langle ij \rangle$ indicates
nearest-neighbour pairs; $\delta$ is the Kronecker delta; and
$\bf{s}^{Q}$ is a fixed direction amongst the $q$ possible ones
(one says that the field $H$ is coupled to the spin-state $Q$).
The model can be modified to include multi-site interactions.

Using its representation in terms of a graph with coloured
edges, Potts model can be generalized to {\em arbitrary,
non-integer values of $q$}. It is connected with  other
interesting models of statistical mechanics. Potts model with
$q=4$ is known as Ashkin-Teller model, Ref.~\refcite{ashkin43};
the case $q=2$ corresponds to the usual Ising model; $q=1$ model
is in  correspondence with the percolation process; $q=1/2$ can be
related to a dilute spin glass model; finally, the $q=0$ limit
relates to the Kirchhoff resistor network problem.

The nature of the order-disorder phase transition in Potts model
depends -- besides on the ferromanetic or antiferromagnetic nature
of the model -- both on the dimension $d$ and on the value of $q$.
For example, for the ferromagnetic model in the absence of a
magnetic field, for $d=2$ there is a second-order phase transition
when $q=2,\,3,\,4$, while for $q>4$ the transition is first-order.
There exists a dimension-dependent critical value of $q$,
$q_c(d)$, above which the transition is mean-field like  -- either
of the first order if $q>\mbox{max}(2, q_c)$, or continuous if
$q_c \leqslant q \leqslant 2$. The few exactly-known points are
$q_c(2)=4$ (Ashkin-Teller), $q_c(4)=2$ (Ising), $q_c(6)=1$
(percolation). Besides the $q=2$ (Ising) case, the $q=3$ and $q=4$
models have a rather well-studied phase diagram, which proved to
be dimension-dependent.

The location and characteristics of the phase transition(s) --
that are known analytically only in few cases -- can be studied,
essentially numerically, in the Yang-Lee formalism. The canonical
partition function, in the general case, is  a function of three
control parameters: the magnetic field intensity $H$, the
temperature $T$, and the parameter $q$ of the model. One can
therefore consider the complex extension of the partition function
and the corresponding zeros either with respect to the complex
magnetic field (Yang-Lee zeros), or the complex temperature
(Fisher zeros), or, moreover, with respect to the complex $q$ --
the so-called {\em Potts zeros}.

\subsection{Yang-Lee zeros}
\label{pottsyl}

A first category of studies were done on the zeros of the partition function in the
plane of the complex magnetic field, at fixed temperature and real value of the
parameter $q$ of the Potts model, for various dimensions $d$ and types of lattices.

The one-dimensional lattice was studied  in the  references cited
below using a transfer-matrix technique. It was shown\cite{kim00I}
that for $1<q<2$ the zeros lie inside the unit circle in the
complex fugacity plane, while for $q>2$ they lie outside the unit
circle for finite temperature; $q=2$ corresponds to the Circle
Theorem for the Ising system. The situation $q<1$ is somehow
pathological\cite{glumac94,kim04II}, since for all the
temperatures the zeros lie (in part or completely, depending on
$T$) on the real axis, and accumulate around a certain point of
the real axis. This would mean that one obtains a second-order
phase transition at finite temperature, and in the presence of a
real magnetic field; however, the eigenvalues of the transfer
matrix -- which give the correlation length -- are not real. $q=1$
is a particular limiting case. The discussion of a tricritical
point and its connection with a special type of Yang-Lee edge
singularity is given in Ref.~\refcite{mittag84} for $q=3$ Potts
model. Reference~\refcite{dolan02} demonstrates a connection
between the locus of the Yang-Lee zeros and the Julia set of the
logistic map associated with the renormalization transformation of
the lattice (and idea already present in the study of Ising models
on hierachical lattices, see above). Finally,
Ref.~\refcite{ghulghazaryan03} presents a recursive method
(adapted from the theory of dynamical systems) for the computation
of the partition function zeros.

The two- and three-dimensional models on regular lattices  are
discussed in  Refs.~\refcite{kim98I,monroe99,kim00I}, and they are
shown\cite{kim00I} to have the same type of behavior with respect
to $q$ as the one-dimensional lattices. Yang-Lee zeros on
recursive Bethe  lattices are studied in
Ref.~\refcite{ghulghazaryan02} for noninteger values of $q$.

\subsection{Fisher zeros}
\label{pottsfisher}

As already mentioned in Sec.~\ref{fisher}, the Fisher zeros are
much more sensitive than the Yang-Lee zeros to the details of
the interactions (type of lattice, dimensionnality), and also to
finite-size effects (like boundary conditions), and this explains
the large number of (essentially numerical) papers devoted to the
study of various specific models.

For the one-dimensional lattices with arbitrary $q$ it was shown
in Ref.~\refcite{glumac94} that the zeros in the complex field
plane can be related, through a kind of duality transformation,
to the zeros in the complex temperature plane. Potts models on
square lattices are discussed, e.g., in
Refs.~\refcite{chen96,matveev96,kim00II}, in the absence of a
magnetic field and for various values of $q$. In the region where
${\cal R}e[\exp(\varepsilon/k_BT)]<1$, Fisher zeros lie on the
circle $|\exp(\varepsilon/k_BT)-1|=\sqrt{q}$, while outside this
region the zeros are no longer ocuppying this circle and lead to
the existence of (nonphysical) points at which the magnetization
and susceptibilities diverge -- the {\em Fisher edges}. The
problem of the Fisher edges is also discussed in
Refs.~\refcite{kim98II,kim01,kim02,kim04III,ghulghazaryan03}, both
in the absence and presence of an external magnetic field.

Finally, a large body of literature addresses these problems on
various other lattices -- strips of square
lattices\cite{chang01,chang02}, honeycomb, triangular, and
Kagom\'e lattices\cite{feldmann98I,feldmann98II,chang04}, and
recursive Bethe lattices.\cite{ghulghazaryan02,dolan01}

\subsection{Potts zeros}
\label{pottspotts}

As already mentioned, Potts model can be extended to arbitrary,
continuous values of $q$ (through the so-called
Kasteleyn-Fortuin\cite{fortuin72} representation). Thus the
canonical partition function becomes a function  of the parameter
$q$, and this gives the motivation for the analysis of its zeros
in the plane of complex $q$ -- the {\em Potts zeros}. The physical
relevance of such an approach may be somewhat unclear at  first
sight. However, in Ref.~\refcite{kim01} it is shown that one can
find a similarity between the scaling property of complex-$q$
zeros and the den Nijs\cite{nijs79} expression for the thermal
critical exponent; in Ref.~\refcite{kim04III} it is shown that a
Circle Theorem holds for a certain range of intensity of the
applied magnetic field, while for other ranges the Potts zeros lie
on the positive real axis (without necessarily being accumulation,
i.e., phase-transition points). A {\em Potts edge} exists,
analogous to the Yang-Lee and Fisher edges. Several types of
lattices are considered in
Refs.~\refcite{ghulghazaryan03,chang01,chang02,chang04}. Finally,
$1D$ long-range interaction and the mean-field cases are
approached in Ref.~\refcite{glumac02},
with a discussion of the phase transition point and of finite-size effects.\\

While most of the above-cited references address the case
of the ferromagnetic Potts model, there is a non-negligible amount of work dedicated
to the {\em antiferromagnetic} case, see Refs.~\refcite{kim00II,chang01,chang04,kim04III,kim01,kim04II}.
In particular, the canonical partition function in the $T\rightarrow 0$ limit reduces to
chromatic polynomials in $q$, that have a relevance in the theory of graphs, Ref.~\refcite{salas01}.\\


\section*{PART II: NONEQUILIBRIUM STEADY-STATE PHASE TRANSITIONS\vspace{0.5cm}\\}
\label{nonequilibriumtransitions}

\section{Generalities}
\label{generalities}

The study of nonequilibrium systems reveals its importance once
one realizes that equilibrium in nature is merely an exception
rather than the rule, and that most of the qualitative, structural
changes (like, for example, pattern formation) that one encounters
in various systems take place under nonequilibrium conditions.

From a macroscopic point of view, roughly speaking, a
nonequilibrium system is the set of fluxes of various
characteristic quantities (e.g., number of particles, energy,
etc.) inside the system and between the system and its
surroundings. Let us try to give a more precise, quantitative
meaning of the notion of nonequilibrium system from a
microscopic point of view, in the frame of a stochastic
description of the system in the corresponding configuration
space. Of course, as for equilibrium systems, this probabilistic
description is imposed by the huge number of microscopic degrees
of freedom, and by the practical necessity of describing the
system in terms of a few relevant measurable quantities.

A configuration $\omega$ of the system represents a set of
relevant mesoscopic variables, corresponding to the degree of
coarse-graining of the adopted description. The ensemble of the
accessible configurations, according to the constraints imposed on
the system, form the corresponding configuration space. The system
is  jumping  between the configurations according to some
prescribed stochastic transition rules. The state of the system at
a time $t$ is given by the probability distribution $P(\omega, t)$
associated to the set of possible configurations $\{ \omega \}$.
Suppose now that the level of coarse-graining is such that the
evolution of the system is Markovian (this is the only type of
systems that we shall be dealing with in the foregoing). The
evolution of the probability distribution function can be
described by a master equation with a set of  transition rates
between the configurations, $\{{\cal W}(\omega \rightarrow
\omega')\}$:\footnote{One can consider the more general case when
these transition rates are time-dependent but, for simplicity,
here we shall limit ourselves to the case of constant transition
rates.}
\begin{equation}
\frac{dP(\omega,t)}{dt}=\sum_{\omega'\neq \omega} \left[\frac{}{}
{\cal W}(\omega' \rightarrow \omega)
P(\omega',t)-  {\cal W}(\omega \rightarrow \omega')
P(\omega,t) \right]\,.
\label{evolprob}
\end{equation}

Let us consider the long-time behavior of the system. The dynamics
can be such that no stationary state is reached (e.g., the system
may exhibit a limit-cycle or a chaotic-like behaviour). On the
contrary, suppose that the system reaches a {\em stationary
state}, i.e., the probability distribution of the possible
configurations becomes time-independent, $P_s(\omega)$. The
stationarity condition in the configuration space is that the
total flow of probability into configuration $\omega$ is balanced
by the corresponding outgoing  flow,
\begin{equation}
\sum_{\omega' \neq \omega} \left[\frac{}{}{\cal W}(\omega' \rightarrow \omega)
P_s(\omega')-  {\cal W}(\omega \rightarrow \omega')P_s(\omega)\right]  =0\,\quad \mbox{for  all} \;\,\omega\,.
\label{stationary}
\end{equation}

One realizes that an equilibrium state is a very particular case
of stationary state. Besides the independence on time of the
characteristic quantities of the system, there is no macroscopic
exchange between the system and its surroundings, i.e., no flow
runs through the system and its borders. From a stochastic point
of view, this corresponds to the {\em detailed-balance} condition
in the configuration space,
\begin{equation}
{\cal W}(\omega' \rightarrow \omega)
P_s(\omega')= {\cal W}(\omega \rightarrow \omega')P_s(\omega)\,\quad
\mbox{for  all} \;\,\omega,\,\omega',\,\omega\neq \omega'\,,
\end{equation}
i.e., the balance of probability flow between {\em any pair of configurations}.

On the contrary, a {\em nonequilibrium stationary state} corresponds to {\em probability
flow loops} in the configuration space, i.e., to the {\em breaking of detailed balance}
for certain configurations $\omega,\,\omega'$,
\begin{equation}
{\cal W}(\omega' \rightarrow \omega)
P_s(\omega')\neq {\cal W}(\omega \rightarrow \omega')P_s(\omega)\,.
\end{equation}
These probability flow loops lead to the observed flows of
macroscopic quantities through the system and its borders. In
particular, note that in order to maintain a system in a
nonequilibrium state continuous exchanges with its surroundings
are necessary, i.e., the system is necessarily {\em opened}.

One should realize that for such stochastic models the physics
is embedded in the transition rates, that are chosen on
``reasonable" backgrounds according to the nonequilibrium phenomena
one is wishing to describe.  It seems thus that one has a high
degree of freedom in the choice of the transition rates that lead
to a nonequilibrium  stationary states, although it has been
argued recently\cite{evans04I,evans04II,evans05} (on the basis of
Jaynes' MaxEnt principle extended to nonequilibrium situations)
that there are  rather severe restrictions on this choice.

To sum-up, {\em breaking of detailed balance is the general
characteristic of stochastic systems in a nonequilibrium
steady-state}.

\section{A partition function  for  nonequilibrium stationary states}
\label{noneqpartition}

Phase transitions in nonequilibrium steady states are sometimes
accompanied, as in the equilibrium case, by a {\em breaking of
ergodicity} of the system, i.e., by the fact that the asymptotic
stationary state of the system is not uniquely-defined, but
depends on its  initial configuration. This means that
Eq.~(\ref{stationary}), for certain values of the transition rates
(depending on the control parameters of the system), admits more
than one solutions for the stationary probability distribution
$P_s(\omega)$, and, correspondingly, the stationary properties of
the system (e.g., the mean values of the characteristic parameters
and of their stationary flows throughout the system)  differ from
one phase to another.

The problem we address now is whether there exist any possibility
to define a partition function  for the nonequilibrium systems,
that allows for the characterization of a nonequilibrium phase
transition between  stationary states in a way that is analogous
to that described in Part I for the equilibrium cases. The answer,
in a rather general frame, was given in Ref.~\refcite{blythe03},
and we present here the relevant arguments, see also
Refs.~\refcite{evans00,blythe01,evans02,gaveau98}.

For simplicity,  define the {\em transition
matrix}\cite{vankampen92} $[W(\omega, \omega')]$, whose
off-diagonal terms are equal to the transition rates
$W(\omega,\omega')={\cal W}(\omega' \rightarrow \omega)$ and the
diagonal terms are $W(\omega,\omega)=-\sum_{\omega'} {\cal W}
(\omega \rightarrow \omega')$, so that the evolution
equation~(\ref{evolprob}) for the probability density  reads
\begin{equation}
\frac{dP(\omega,t)}{dt}=\sum_{\omega'}
W(\omega,\omega') \,P(\omega', t)\,,
\label{evolprobagain}
\end{equation}
and it preserves the normalization condition $\sum_{\omega} P(\omega, t)=1$.
The stationarity condition (\ref{stationary}) reads
\begin{equation}
\sum_{\omega'}
W(\omega,\omega') \,P_s(\omega')=0\,, \quad \mbox {for all}\; \omega\,.
\label{stationaryagain}
\end{equation}

Suppose now that the control parameters of the system are such
that there is a unique steady state. Then, according to the
Perron-Frobenius theorem, there exists a single eigenvalue of the
transition matrix that is equal to zero, and the corresponding
eigenvector gives the stationary-state probabilities of the
various configurations $P_s(\omega)$. All the other eigenvalues
$\{\lambda_i\}$ of the transition matrix have negative real parts,
${\cal R}e \lambda_i >0$, and each of them corresponds to an
eigenvector of the transition matrix that relaxes exponentially to
zero within a time-scale of the order $1/|{\cal R}e \lambda_i|$.

Because of the zero eigenvalue of the transition matrix, the
linear set of equations (\ref{stationaryagain}) is
underdetermined, and therefore one can obtain from it only  the
{\em relative statistical weights} $f_s(\omega)$ of the various
configurations. However, one can define the {{\em normalization
factor}
\begin{equation}
Z=\sum_{\omega} f_s(\omega)\,,
\end{equation}
such that
\begin{equation}
P_s(\omega)=\frac{1}{Z}\,f_s(\omega)\,.
\end{equation}
This normalization factor is related to the nonzero eigenvalues of
the transition matrix, more precisely (up to a multiplicative
factor),\footnote{In order to obtain this result, note that the
characteristic polynomial of the transition matrix reads (up to a
multiplicative factor)
$$
\mbox{det}(\lambda I-W)= \left[\mbox{det} (W)+ \lambda \sum_{\omega} f_s (\omega) + {\cal O} (\lambda^2) \right]\,.
$$
Given that $\mbox{det}(W)=0$, and $Z=\sum_{\omega} f_s (\omega)$, one has:
$$
Z=\lim_{\lambda \rightarrow 0} \frac{\mbox{det}(\lambda I-W)}{\lambda}=\lim_{\lambda \rightarrow 0}\frac{1}{\lambda}\displaystyle\prod_{\mbox{all}\, \lambda_i}(\lambda-\lambda_i) = \prod_{\lambda_i \neq 0}(- \lambda_i)\,.
$$ }
\begin{equation}
Z=\prod_{\lambda_i \neq 0} (-\lambda_i)\,.
\label{basic}
\end{equation}
This is the key relationship that allows to propose {\em $Z$ as a
candidate for a nonequilibrium steady-state partition function}
that can give informations on the appearance of a phase transition
in the system.

Indeed, suppose that one varies the control parameters (and thus
the transition rates, as well as the corresponding eigenvalues of
the transition matrix) so that the system is driven towards a
nonequilibrium phase transition. A new stationary state appears in
the system, and, correspondingly,  at a phase transition one is
expecting the appearance of a diverging time scale related to this
new slow-mode that sets in. This means that one of the eigenvalues
$\lambda_i$ that is associated to this  mode goes to zero as one
approaches the phase transition point and thus, by virtue of
Eq.~(\ref{basic}), so does the normalization factor $Z$.

It seems therefore meaningful to try to locate the nonequilibrium
phase transition points through the analysis of the zeros of the
normalization factor $Z$. However, the legitimacy of this approach
has to be checked on several known exactly-solvable models before
trying to use it for further predictions. Several classes of
nonequilibrium systems were discussed in the litterature using
this approach, and we briefly review them here, following for the
most the original work of Blythe  and Evans,
Ref.~\refcite{blythe03}.

Recall at this point that $Z$ is defined up to a multiplicative
factor,\footnote{Note that such a spurious multiplicative factor
may appear also in equilibrium systems, e.g., through a uniform
shift of the energy scale.} that depends on the method adopted to
solve Eqs.~(\ref{stationaryagain}); this factor (which is commoun
to all the stationary states weights $f_s$) is generally  a
polynomial in the transition rates ${\cal W}(\omega \rightarrow
\omega')$, and therefore may introduce additional zeros to $Z$.
However, one expects that these zeros are not physically relevant,
and this can be checked on the models that are discussed below.

\section{Driven diffusive systems}

Most of the studies on the applicability of Yang-Lee theory to
nonequilibrium phase transitions refer to simple models of systems
with diffusion and drift, mostly because they were intensively
studied in the last years and many results are known for their
stationary states and phase diagrams.

\subsection{Stochastic single-species models}
\label{TASEP}

Such models belong to the class of the so-called ASEP (for
``Asymmetric Simple Exclusion Process"). These are
interacting-particle systems, with a single species of particles
with no internal degrees of freedom, and with hard-core two-body
interactions with their nearest-neighbour sites on a
one-dimensional lattice. This hard-core interaction with excluded
volume is expressed by the exclusion condition that each lattice
site may be occupied by at most one particle. Therefore this class
of models may be described by a set of occupation numbers $\{n_1,
n_2, ... , n_L\}$, where $n_k=0,\,1$ is the number of particles on
site $k$ of a lattice of $L$ sites. The particles may jump at
random, according to some prescribed stochastic rules, by one
lattice site, with a bias in the right/left transition
probabilities. The system is maintained in a nonequilibrium
stationary state, with a constant particle flow along it, by
continuous injection and extraction of particles at its two
borders. The specificity of each model is embedded in the
prescribed stochastic dynamics, i.e., in the probabilities for the
changes of the configuration of a pair of neighbouring sites
$k$ and $k+1$, and also in the continuous or discrete (parallel or sequential) character of the dynamics, see, e.g., Refs.~\refcite{schmittmann95,schutz01} for an overview. \\

The  {\em totally asymmetric exclusion process} (TASEP) with open
boundaries is perhaps the simplest exactly solvable model with a
nontrivial stationary behavior that includes both a first-order
and continuous phase transitions, and, in view of its prototypical
character, we will describe it here in some detail, see also
Refs.~\refcite{blythe02,blythe03}.

In its {\em continuos-time} variant, in an infinitesimal time
interval $dt$, one can have one of the following transitions: a
particle may hop to the right with probability $dt$ (i.e., with
transition rate equal to unity), provided the receiving site is
empty; or a particle is injected at the left border (leftmost
lattice site) with probability $\alpha\, dt$ (provided the left
border is empty); or a particle at the right border is removed
from the system with probability $\beta\, dt$ (provided the right
border contains a particle). Obviously, the injection and
extraction rates, $\alpha$ and $\beta$, are the control parameters
that determine the stationary state of the system, that can be
characterized through the density profile, i.e., the set of mean
occupation numbers of the lattice site, $\langle n_k \rangle,\,
k=1, ... , L$. The stationary flow of particles $J=J(\alpha,
\beta)$ exhibits non-analiticities  {\em in the thermodynamic
limit} of an infinite lattice $L \rightarrow \infty$, as one
varies $\alpha$ and $\beta$, and these non-analiticities
correspond to phase transitions in the system.
 The phase diagram consists of three regions: (i) A high-density
phase with a density profile that decays exponentially towards the
right boundary. In this phase the current $J=\beta(1-\beta)$ is
controlled by the  rate at which the particles are removed from
the system, $\beta < \mbox{min}(\alpha, 1/2)$. (ii) A low density
phase, with a density profile that has an exponential decay
towards the left boundary. The current $J=\alpha(1-\alpha)$ is
again controlled by the smallest of the two rates, $\alpha <
\mbox{min}(\beta, 1/2)$.  At the first-order transition line
$\alpha=\beta<1/2$, the current exhibits a discontinuity in its
first derivative. The system presents a {\em shock} separating
regions of high and low densities, and this is an example of {\em
phase coexistence} at a nonequilibrium first-order phase
transition. (iii) The third phase, for $\alpha, \beta >1/2$, is
called the maximum current phase, since the current assumes
throughout the constant value $J=1/4$, which is the largest
possible current for any combination of $\alpha$ and $\beta$. The
density profile decays {\em as a power-law} from both boundaries,
and therefore the transitions from either phase (i) or phase (ii)
are accompanied by diverging lengthscales. Moreover, $J$ has a
discontinuity in its second derivative, and this corresponds to a
second-order phase transition.

The normalization factor for the TASEP is
\begin{equation}
Z_L(\alpha,\beta)=\sum_{\ell=1}^L \displaystyle \frac{\ell (2L-1-\ell)!}{L! (L-\ell)!}\;
\displaystyle\frac{(1/\beta)^{\ell+1}-(1/\alpha)^{\ell+1}}{1/\beta-1/\alpha}\,,
\end{equation}
and the current $J=Z_{L-1}/Z_L$. In the thermodynamic limit  $\ln
Z_L \approx -L \ln J$, i.e., $\ln J$ plays the role of a free
energy density for the nonequilibrium partition function
$Z_L$. From here it is routine to apply the Yang-Lee procedure to
$Z_L$. For example, keeping $\beta$ as a parameter, one finds that
the zeros of $Z_L$ in the plane of the complex variable $\alpha$
lie on a smooth curve, and accumulate at the phase transition
point $\alpha_c$ on the real positive semiaxis when $L$ is
increased towards the thermodynamic limit. Depending on the value
of $\beta$, one has: (i) If $\beta <1/2$, the curve of zeros
passes smoothly, with a nonzero density of zeros, through the
transition point $\alpha_c=\beta$, corresponding to a first-order
phase transition, see Sec.~\ref{generalframe}; (ii) If $\beta >
1/2$, the density of zeros decays to zero as one approaches the
transition point $\alpha_c=1/2$, and the zero-lines approach the
transition point at an angle $\pi/4$ to the real axis (meeting at
a right angle), thus indicating a second-order phase transition.

Therefore the  example of TASEP seems  to indicate that it is
legitimate to study the zeros of the normalization factor $Z$, in
the plane of one complexified control parameter and  in the
thermodynamic limit, in order to get the location of a
nonequilibrium phase transition point. Moreover, the distribution
of these zeros offers informations on the characteristics of the
transition -- exactly as in the equilibrium case.

The same conclusions hold for the time-continuous {\em partially
asymmetric exclusion process} (PASEP), for which the particles
have a nonzero probability to jump to the left, corresponding to a
transition rate $q$. The PASEP has thus a supplementary control
parameter $q$, and one can follow the change in the locus of the
partition function zeros in the complex-$\alpha$ plane (at fixed
value of $\beta$) with changing $q$. Also, one could look for the
zeros of $Z$ in terms of complex $q$ (at fixed values of $\alpha$
and $\beta$). In particular, at $q=1$ (when there is no bias
between right and left transitions), one transits to  a regime of
zero-current in the thermodynamic limit, and this phase transition
should manifest as an accumulation of the complex-$q$ zeros in the
vicinity of $q=1$, a scenario that has still to be verified (see
Ref.~\refcite{brak04I} for some more comments on this phase
transition).

Two types of {\em discrete-time} variants of TASEP were discussed recently in the
frame of Yang-Lee formalism for phase transitions, namely:\\
(i) The {\em parallel-update} TASEP, \footnote{This is a special
case of the famous Nagel-Schreckenberg model for traffic flow [K.
Nagel and M. Schreckenberg, {\em J. Phys. I. France} {\bf 2}, 2221
(1992)], in which the TASEP phase transitions can be considered as
jamming transitions.}  see Ref.~\refcite{blythe04I}. At each time
step particles are introduced with a probability $\alpha$ at the
left border (if this first site is empty) and are extracted with
probability $\beta$ at the right border (if this last site is
occupied); moreover, the particles in the bulk can jump to the
right with probability $p$ (if the right space is empty), or
remain still with probability $1-p$. This update is applied to all
particles {\em simultaneously}. Here again one can compute exactly
the normalization factor $Z=Z(\alpha,\beta,p)$, and look for its
zeros in the complex-$\alpha$ plane at fixed $\beta$ and $p$. The
phase diagram is similar to that of the continuous-time TASEP;
there are three phases,  separated, respectively, by two
continuous-transition lines at $\alpha$ or $\beta = 1-\sqrt{1-p}$,
and a first-order transition line at $\alpha=\beta<1-\sqrt{1-p}$.
Thus the zeros of $Z$ in the complex-$\alpha$ plane also behave
similarly. One can also look at zeros of $Z$ in the complex-$p$
plane (at fixed values of $\alpha$ and $\beta$); their locus is a
cardioid, with the transition point appearing as a cusp on the
real
positive semiaxis.\\
(ii) The {\em sublattice-parallel update} TASEP with the
supplementary constraint of fixed average density of particles in
the system, $\rho=M/L$ (where $M$ is the total number of particles
on the lattice of length $L$) was studied in
Ref.~\refcite{jafarpour05}. In this model, at the first time step,
besides the injection and extraction of the particles at the
borders with probabilities $\alpha$ and $\beta$, respectively, all
the particles on odd sites jump to the right with probability one
(if the site to the right is empty); at the next step, all the
particles on even sites jump to the right with probability one
(again, if the receiving site is empty). Then one repeats this
two-stage process. The phase diagram of this model in the
thermodynamic limit contains three phases, namely a high-density
phase and a low-density phase (as for the continuous-time TASEP),
and a {\em shock phase}, characterized by existence of shocks  in
the particle-density profile. This shock phase has second-order
transition borders with the high and low density phases, and this
is confirmed by the behavior of the zeros of $Z$ in the
complex-$\alpha$ plane, for various fixed values of $\beta$ and
$\rho$.

The problem of finite-size scaling and universality for several
variants of {\em discrete-time} TASEP was addressed recently in
Ref.~\refcite{brankov05}. In the thermodynamic limit, in general
(i.e., whatever the adopted update rule), both first and second
order boundary-induced transitions between the steady-states are
encountered, and one derives finite-size scaling expressions for
the current $J$, for the mean local particle density, as well as
for the zeros of the normalization factor $Z$ in the
complex-$\alpha$ plane. In particular, simulations suggest that
the smallest distance between the locus of these zeros and the
critical point $\alpha_c$ has a power-law behavior  $\sim
L^{-1/\nu}$, where   $\nu$ (as in the equilibrium systems)   is
the critical exponent of the bulk correlation length for the
nonequilibrium phase transition.

\subsection{Two-species models}
\label{twospecies}

Actually, there are two slightly different variants of the same
model that were studied, from the point of view of the Yang-Lee
approach, in Refs.~\refcite{arndt00,jafarpour03I}. As described in
Ref.~\refcite{arndt00}, the model consists of two types of
particles, $A$ and $B$, of fixed numbers $N_A$ and $N_B$, that
diffuse in opposite directions on a one-dimensional $L$-site
lattice with periodic boundary conditions. Specifically, particles
of type $A$ jump to the right  and particles of type $B$ jump to
the left, both with unit transition rates, provided that the
destination site is empty. Also, if on neighbour sites, the $A$
and $B$ particles exchange places, $A+B\rightarrow B+A$ with
transition rate $q$, and  $B+A\rightarrow A+B$ with unit
transition rate. The bias parameter $q$ is thus the control
parameter of the system (at fixed particle densities
$\rho_{A,B}=N_{A,B}/L$), and as long as $q\neq 1$ the system can
reach a nonequilibrium stationary state with a stationary nonzero
flow of particles throughout the system. Arndt\cite{arndt00}
studied the phase diagram in the particular case of equal particle
densities $\rho_A=\rho_B\equiv \rho=M/L$ in terms of the Yang-Lee
formalism. Since the normalization factor$Z_L(q,\rho)$  was not
know analytically at that moment, Arndt went to the numerical
study of a grand-canonical partition function
\begin{equation}
\Xi_L(z,q)=\sum_{M=0}^{M_{max}=L/2}z^M Z_L(q,M/L)
\end{equation}
in the plane of the complex fugacity $z$. From these studies of
the zeros on finite-size systems (with $L$ up to $100$), he
inferred  the existence, in the thermodynamic limit,  of a
first-order phase transition at some point $q_c(\rho)>1$, between
a mixed phase (corresponding to a condensate made of both $A$
and $B$ particles; a block of empty sites, with a few residual
$A$ or $B$ particles, occupies the rest of the system) and a
disordered phase (with uniform density profiles of both species
and no spatial condensation).

 However, it was shown later, see
 Refs.~\refcite{blythe03,rajewsky00,kafri02}, through an exact
 calculation of the grand-canonical partition function in the
 thermodynamic limit, that no phase transition stricto sensu
 appears in the system. When one varies the control parameters,
 there appears, instead, a very abrupt increase in the correlation
 length to an anomalously large  value  -- of the order ${\cal
 O}(10^{70})$ sites, which, however, if finite. Although very
 spectacular, this phenomenon does not seem to be extremely rare,
 since it can be argued, see, e.g., Ref.~\refcite{kafri02}, that
 it is generated when the dynamics of the model is such that small
 domains tend to be suppressed in the steady-state distribution.

Thus the extrapolation done by Arndt from numerical results on
finite-size systems towards the thermodynamic limit is unreliable.
One should therefore be aware of the possible appearance of such
extremely large correlation lengths in the nonequilibrium systems,
correlations that may lead to a false impression of the existence
of a phase transition when investigating (numerically) systems
with sizes less or comparable to the correlation length.

Note also\cite{blythe03} that investigating the zeros of this
grand-canonical partition function in the complex fugacity plane
amounts, from a physical point of view, at placing the lattice in
equilibrium with a reservoir of particles, and thus (despite the
claim in the title of Ref.~\refcite{arndt00}) at studying an
equilibrium system.

Reference~\refcite{jafarpour03I} reconsidered the same model in
the exactly-solvable case $N_B=1$, i.e., of a single impurity
present in the system. Using a matrix-product formalism, one can
compute the corresponding normalization factr $Z_L(q,\rho)$ at
fixed density $\rho$ of the $A$ particles. Studying the behaviour
of $Z_L(q,\rho)$ in the  thermodynamic limit, one deduces the
existence of two stationary phases of the system, namely, (i) a
{\em jammed phase} for $q<2\rho$, when the impurity provokes a
macroscopic shock (error-function like) in the density  profile of
$A$ particles; (ii)  and a {\em power-law phase} for $q>2\rho$,
when the impurity has a short-range effect on the system, such
that the density profile of $A$-s has an exponential behaviour,
with a correlation length $\sim \left|\ln(q_c/q)\right|^{-1}$ that
diverges when $q$ approaches the critical value $q_c=2\rho$.  This
result is confirmed by the study of the location and distribution
of zeros of $Z_L(q,\rho)$ in the complex-$q$ plane (at fixed value
of $\rho$), that predict correctly the second-order phase
transition point between the two phases, as well as its
characteristics.

\section{Simple reaction-diffusion models}

These stochastic models are characterized by the fact that,
contrary to the drive-diffusive systems discussed above, the
number of particles is not conserved by the dynamics, i.e., there
are some reactions inside the system. One such model, which
results through a simple modification of the continuous-time ASEP,
is the {\em coagulation-decoagulation model} with open boundaries,
see Ref.~\refcite{jafarpour03II}. It consists of particles that
diffuse, coagulate (two neighbour particles merge into a single
one), and decoagulate (one particle splits into two neighbour
particles) preferentially, e.g.,  in the leftward direction. More
precisely, in the simplified model in Ref.~\refcite{jafarpour04I},
they (i) diffuse to the left and right with rates $q$ and $q^{-1}$
respectively; (ii) coagulate at the left $A+A \rightarrow A+0$
with rate $q$ and at the right  $A+A \rightarrow 0+A$ with rate
$q^{-1}$; (iii) decoagulate to the left $0+A \rightarrow A+A$ with
rate $\gamma q$ and at the right $A+0 \rightarrow A+A$ with rate
$\gamma q^{-1}$. Here $q>1$ to assure the leftward preference of
the processes. Besides this, there is an injection and an
extraction of particles solely at the left border, with rates
$\alpha$ and $\beta$. This model, for which the normalization
factor $Z$ can be computed exactly when
$\alpha=(q^{-1}-q+\beta)\gamma$, exhibits a first-order phase
transition between a low-density and a high-density phase. The
roots of $Z$ in the complex-$q$ plane are studied (for fixed
$\beta$ and $\gamma$), and Yang-Lee theory is shown to hold,
describing correctly both the location and the nature of the phase
transition.

A variant of this model with reflecting boundary conditions,
instead of the open boundaries, was studied in
Ref.~\refcite{jafarpour04I}. The model in this case has two
control parameters, $q$ and $\gamma$, at a fixed particle-density
$\rho$. The system has two second-order phase transition points,
at  $q_c=1/\sqrt{1-\rho}$ and $q'_c=1/q_c=\sqrt{1-\rho}$. The
reason of the existence of these two transition point is simply
the invariance of $Z$ under the transformation $q \rightarrow
q^{-1}$. Note that $\gamma$ plays no role in the location of the
phase transition point. Again the Yang-Lee
formalism predicts correctly these results.\\

Reference~\refcite{jafarpour04II} offered recently a unified
frame, based on the matrix-product formalism, for the study of the
nonequilibrium steady-state phase transitions of three families of
one-dimensional nonequilibrium models with opened boundaries.
Besides the TASEP and a generalization of the above-described
coagulation-decoagulation model, it also discusses the {\em
asymmetric Kawasaki-Glauber process} (AKGP). The latter
corresponds to a (totally) asymmetric diffusion of the particles
(Kawasaki spin-exchange dynamics), combined to death to the left
and right $A+0 \rightarrow 0+0$, $0+A \rightarrow 0+0$, and to
branching to the left and right $A+0 \rightarrow A+A$,
$0+A\rightarrow A+A$. These two last processes represent a variant
of the spin-flip Glauber dynamics. Moreover, a steady-state
current is maintained through the injection and extraction of
particles at the borders. For special values of the corresponding
transition rates, one can estimate the normalization factor $Z$,
and study its Yang-Lee zeros (however, the paper does not
concentrate further on this neither for the AKGP, nor for TASEP,
but only for the extended coagulation-decoagulation model).

\section{Directed percolation}
\label{DP}

An important type of nonequilibrium phase transitions is
represented by the {\em absorbing-state phase transitions}, which
take place when a system, during its evolution, reaches a
configuration (a so-called absorbing state) in which it remains
trapped for ever. Phenomena of this type are present in a wide
variety of physical, chemical, biological systems, see
Refs.~\refcite{droz03,odor04,hinrichsen00} for recent reviews. It
has been conjectured recently that a large group of such models
fall into the {\em directed-percolation (DP) universality class}
(see Sec.~\ref{connection} below for some comments on universality
classes in nonequilibrium systems, and
Refs.~\refcite{grassberger82,janssen81} for this conjecture). A
simple example of such a model is a reaction-diffusion system on a
one-dimensional lattice,  with a competition between the
reproduction of particles, i.e., a (symmetric) decoagulation,
$A+0\rightarrow A+A$ or $0+A\rightarrow A+A$, and the death of
particles, i.e., a spontaneous decay $A\rightarrow 0$. This
competition is controlled by some external parameter $p$; if its
value is below a certain threshold $p<p_c$, the system reaches an
absorbing state (empty lattice) with certainty; otherwise the
system remains active forever in the thermodynamic limit of an
infinite-lattice. The corresponding phase transition is a
continuous one.

In order to study the applicability of the Yang-Lee theory to this
class of nonequilibrium phase transition, it is natural to turn to
the original {\em directed percolation model}, see
Refs.~\refcite{arndt01,dammer01,dammer02,blythe03}.  In order to
fix the ideas, let us consider the two-dimensional case --
although the process can be defined on general $d$-dimensional
lattices. The system consists of a finite square lattice with $L$
rows, that has one site, the apex $O$, on the first row, two sites
on the second row, etc., up to a total of $L(L+1)/2$ sites.  Bonds
between the sites are opened with probability $p$, and
closed with probability $1-p$.  At the initial moment, a
particle is placed on the apex; at each time step this particle
jumps on a site of the  next  lower level, provided the
corresponding bond is opened. The sites connected by a continuous
percolation path (i.e., by a succession of opened bonds) are
said to belong to the same percolation cluster. One is interested
by the {\em percolation (or survival) probability} $P_L(p)$  that
at least one site on the level $L$ is connected to the apex
through a percolation path;  $P_L(p)$ is obtained by considering
all the possible bond configurations. The order parameter,
defined for an infinite lattice $L\rightarrow \infty$, is
$P_{\infty}(p)$, which is the probability of having an infinite
cluster (i.e., a cluster that goes from the apex all the way down
the lattice) for a given bond probability $p$. If $p<p_c$, then
$P_{\infty}(p)$ is zero, i.e., there are no infinite percolation
clusters in the system (that means that one can always, i.e., with
probability one, find a row, sufficiently far from the origin,
that is not connected to the apex through any percolation path.
For $p>p_c$, however, $P_{\infty}(p)$ becomes nonzero, and in the
vicinity of the critical point it has a power-like behavior,
\begin{equation}
P_{\infty}(p) \sim (p-p_c)^{\beta}\,,
\end{equation}
with an universal exponent $\beta$. In addition, the DP process is
characterized by a transversal correlation length $\xi_{\perp}$
perpendicular to the percolation flow,   and a longitudinal
correlation lenght $\xi_{||}$ along the percolation flow, that
diverge both when $p$ approaches its critical value $p_c$,
\begin{equation}
\xi_{\perp} \sim |p-p_c|^{-\nu_{\perp}}\,\quad \xi_{||} \sim |p-p_c|^{-\nu_{||}}\,.
\end{equation}
 A continuous transition takes place at $p_c$. Although DP can be
defined and simulated easily, no analytical solution exists for it
(neither for any of the models pertaining to the DP universality
class) in dimension $d$ lower than the critical dimension $d_c=6$.
Therefore, one has to use numerical estimates of the critical
exponents. Currently, the best estimates\cite{jensen96} in $d=2$
are $\beta=0.27649$, $\nu_{\perp}=1.096854$, and
$\nu_{||}=1.733847$; the percolation threshold depends both on
the spatial dimension and the type of array, and for $d=2$ on a
square lattice the best estimation of it is $p_c=0.6447001$.

Attempts to describe this phase transition in the Yang-Lee
formalism were done in Refs.~\refcite{arndt01,dammer01,dammer02},
by considering the zeros of $P_{L}(p)$ in the plane of complex
$p$, for dimensions $L\leqslant 15$, and looking for their
accumulation points when one increases $L$, i.e., when one goes
towards the thermodynamic limit. These zeros do not lie on a
single smooth curve (as it was the case for the other
nonequilibrium systems discussed above), but on a sequence of
curves that tend to intersect at the critical point. Also, the
results in Refs.~\refcite{arndt01,dammer01,dammer02} suggest a
fractal structure of the distribution of these zeros (but this
aspect deserves further investigation). From these numerical
results, the authors infer the values of  $p_c$, and (using
finite-size scaling arguments) also the longitudinal exponent
$\nu_{||}$.

It should be noted, however, that the partition function $P_{L}(p)$ has some properties that render it qualitatively different from the other nonequilibrium partition functions considered above.
 A first remark~\cite{arndt01,baxter88} is that the survival
probability $P_{L}(p)$ can be represented formally as the
partition function of a system of Ising spins, with two-spin and
three-spin plaquette interactions, on a pyramid-shaped lattice.
However, some of these three-spin interactions have an infinite
energy, and thus the usefulness of such a mapping is questionable.
Also, one encounters some problems with the definition of a
thermodynamic limit and of an extensive free energy of the system,
since, on one hand, $P_L(p)$ does not grow exponentially with the
size of the system, and on the other hand $\lim_{L\rightarrow
\infty}P_L(p)=0$ for $0\leqslant p <p_c$, and it is not clear how
to define a free energy density in this case. This illustrates
some of the difficulties related to the extension of equilibrium
concepts to nonequilibrium situations.

\section{Self-organized criticality}
\label{SOC}

Two recent studies, Refs.~\refcite{cessac02,cessac04}, addressed
the problem of the applicability of the Yang-Lee formalism to
self-organized criticality (SOC). Roughly speaking, a system that
exhibits SOC is a nonequilibrium system (subject to external
stationary nonequilibrium constraints) characterized by the fact
that, in the thermodynamic limit, it reaches a stationary state
with {\em scale invariance and power-law statistics}. One should
underline that this nonequilibrium state is reminiscent of an
equilibrium critical states; however, contrary to these ones, this
stationary critical state is reached solely through the internal
dynamics of the system in response to the external constraints,
without the fine tuning of the control parameters that is
required for the equilibrium criticality. There is a huge body of
literature devoted to this topic, see Ref.~\refcite{jensen98} for
a pedagogical overview of the field.

The papers of Cessac {\em et al.} rise then the legitimate
question whether this spontaneous criticality manifests itself
in the properties of the distributions of complex zeros of a
properly defined partition function for the stationary state,
i.e., if these zeros have an accumulation point when the size of
the system is increased to infinity, without any manipulation
of the parameters of the system.

The dynamics of SOC systems occurs in avalanche-like events;
one can define a set of observables (size, duration, etc.) characterizing these
avalanches. In a finite system of size $L$ such an observable ${\cal V}$
can take a finite set of values, up to a maximum finite value $N_L$,
${\cal V}=0,\,1,...,N_L$, according to a stationary probability distribution
$P_L({\cal V}=n)$. Let us consider the generating function of this distribution
(the grand-canonical partition function),
\begin{equation}
\Xi_L(z)=\sum_{i=1}^{N_L} z^n P_L(n)\,,
\end{equation}
that is a polynomial of degree $N_L$ in the fugacity $z$. If
the system is SOC, the zeros of this polynomial in the complex-$z$
plane should have an accumulation point on the real positive
semiaxis in the thermodynamic limit $L\rightarrow \infty$ for any
parameters of the system. This reasoning was verified in the
above-cited references for various finite-size scaling forms of
$P_L(n)$ encountered in the literature on SOC: the zeros pinch the
real axis at $z=1$ as the system size goes to infinity. A scaling
theory for the Yang-Lee zeros is proposed in this setting, that
shows, under specific conditions, a violation of the scaling
usually observed in equilibrium critical phenomena.

Note however that the remark we  made in Sec.~\ref{twospecies}
with respect to Ref.~\refcite{arndt00} applies here, too, namely
that considering the generating function $\Xi_L(z)$ amounts
actually at analysing an equilibrium problem in a grand-canonical
statistical ensemble.

\section{Connection with equilibrium systems with long-range interactions}
\label{connection}

The examples in the above sections show a surprising analogy
between the equilibrium and the noneqilibrium steady-state phase
transitions, as far as the behaviour of the zeros of the
respective partition function in the plane of the complex control
parameter is concerned. The question that arises is then whether
or not this analogy can be pushed further, e.g., if one can
find a kind of (general rule for a) correspondence between a
nonequilibrium steady-state system and an equilibrium one.

As discussed in Sec.~\ref{generalities}, a nonequilibrium system
is characterized by a breaking of detailed balance in the
configuration space, which corresponds to the appearance of
macroscopic flows throughout the system. This generates {\em
effective long-range interactions} and correlations in the system,
as well as the emergence of an effective criticality in
nonequilibrium steady-states, and, in this logic, of universal
distribution functions for macroscopic quantities, of the
corresponding classes of universality, critical exponents, etc.,
see Refs.~\refcite{racz02,racz03,evans00,odor04} for a discussion
of these fundamental issues.

Therefore, if one  is searching for a kind of correspondence
between nonequilibrium steady-states and equilibrium systems, the
latter ones have to have long-range interaction Hamiltonians. The
history of searching for these effective Hamiltonians (and
related problems like, e.g., definitions of an effective nonequilibrium
temperature) is long and tortuous and will not be
discussed here. In particular, trying to write down an explicit
expression of this Hamiltonian does not seem to be a very fruitful
approach.

Here we will  briefly present a few recent results in this
direction, see
Refs.~\refcite{brak04I,brak04II,blythe04I,blythe04II}. These
studies refer to ASEP-like systems, but  in our opinion, in view
of their  generality, these ideas carry an important potential for
further applications. The adopted strategy relies on two
key-elements as discussed
below.\\

\noindent{\bf (A) The formal definition of ``particle numbers" and
associated ``fugacities"}. Let us consider a Markovian system as
described in Sections~\ref{generalities}, \ref{noneqpartition},
and the corresponding steady-state normalization factor $Z_n(\{
{\cal W}(\omega \rightarrow \omega') \})$ (here $n$ designates the
number of configurations $\omega$ accessible to the system; it is
also  explicitly indicated that the normalization factor is a
function of the transition rates). A first point discussed in
Ref.~\refcite{brak04I} is that the normalization factor is a {\em
polynomial in the transition rates ${\cal W}(\omega \rightarrow
\omega')$, with positive coefficients}, of degree $n-1$, i.e., it
has the form of a {\em generating function}.  By the
Cauchy-Schwartz inequality it follows that its negative logarithm
(that we would like to assimilate with a free energy of the
system),
\begin{equation}
F_n=-\ln(Z_n)\,,
\end{equation}
is a {\em convex function} in all its arguments ${\cal W}(\omega
\rightarrow \omega')$. One is then tempted, by analogy with
equilibrium situations, to identify {\em formally}  the transition
rates with ``fugacities", and to consider the corresponding
``particle numbers"
\begin{equation}
N_{\omega,\omega'}=-\,[{\cal W}(\omega \rightarrow \omega')]\,
\frac{\partial Z_n}{\partial [{\cal W}(\omega \rightarrow \omega')]}\,.
\end{equation}
These are well-behaved thermodynamic quantities, in the sense
that they are positive and increasing functions of the fugacities
for any size $n$ of the system. They are, however,
linearly-dependent, and some of them may even coincide (because
some of the transition rates may be zero or may be equal between
them). In the thermodynamic limit of large $n$
\begin{equation}
N_{\omega,\omega'}=V(n)\rho_{\omega,\omega'}\,,
\end{equation}
where $V(n)$ is the ``volume"
(defined by the leading asymptotic behaviour of the normalization)  and $\rho_{\omega,\omega'}$  are the ``densities".

One has now all the {\em formal} ingredients to discuss a
nonequilibrium steady-state phase transition by analogy with the
equilibrium situations. For example, a first-order phase
transition would correspond to a discontinuity of the fugacity as
a function of the density. It is not clear however, in general,
that all the known characteristics of the phase transitions (e.g.,
diverging correlation length for  continuous one) are recovered
within this formalism.

There is one more very important point to be made, and which
represents a major difference between nonequilibrium systems and
equilibrium systems with short-range interactions. It is the fact
that the ``particle numbers" defined above are not necessarily
extensive quantities. This implies that in the space of the
parameters of the system (which are the transition rates), besides
the regions where the ``particle densities" are well-behaved and
finite, there may appear regions where $\rho_{\omega,\omega'}$ may
diverge (and one has to change then the definition of $V(n)$). The
frontiers between these regions correspond then to phase transitions
that do not have a correspondent in the short-ranged interactions
equilibrium systems. Note, however, that phase transitions of this
type are know in {\em equilibrium systems with long-range (or
nonlocal) interactions}\cite{brak04I}, a remark that goes in the
sense of the general statements made
in the beginning of this paragraph. \\

\noindent{\bf (B) The correspondence with equilibrium systems}.
Once this formal frame is set-up, comes the most delicate part,
which is to find a way (possibly a systematic one) to assign a
physical meaning to the normalization factor, ``particle
densities", etc. in a properly-defined {\em equilibrium}
statistical problem. The general solution proposed in
Ref.~\refcite{brak04I}, and illustrated on concrete examples in
Refs.~\refcite{brak04I,brak04II,blythe04I,blythe04II} comes from
the combinatorial graph theory, and consists in relating
explicitely the normalization of a stationary state to the
combinatorial problem of counting weighted spanning trees on
graphs; the weights of the configurations depend on parameters
which correspond to the transition rates of the nonequilibrium
stochastic process. This implies directly an interpretation of the
normalization factor as a statistical mechanics partition sum. Of
course, establishing the right correspondence is a difficult
task, and no general recipe can be given for it.

A first heuristic application of this approach is given in
Ref.~\refcite{degier04}, where a one-dimensional,  stochastic,
adsorption-desorption nonequilibrium model for interface growth
(the raise-and-peel model) was put in correspondence with a
two-dimensional ice model with domain-wall boundary conditions (an
equilibrium problem with nonlocal interactions). Brak and
Essam\cite{brak04II} showed that the matrix representation of the
stationary-state algebra of TASEP can be interpreted
combinatorially as various weighted lattice paths. The
normalization factor of TASEP is identical to the equilibrium
configuration sum of a polymer chain having a two-parameter
interaction with a surface (the so-called one-transit walk). As
shown in Ref.~\refcite{brak04I}, the TASEP current and density are
simply related to the equilibrium densities of the latter model;
in particular, the second-order phase transitions of the TASEP
(see Sec.~\ref{TASEP}) are related to a special type of surface
phase transition encountered in polymer physics. Later on Blythe
{\em et al}~\cite{blythe04I} found that the normalization factor of
the discrete-time parallel-update TASEP can be expressed as one of
several equivalent variants of  two-dimensional lattice
constrained-path problems. Finally, in Ref.~\refcite{blythe04II}
it was shown that the grand-canonical normalization factor of
the TASEP (that corresponds to the immersion of the lattice in a
reservoir of particles, with fixed fugacity $z$) can be computed
in a closed analytical form, and thus it allows a direct
derivation of the asymptotics of the canonical normalization
for the various phases, and the immediate correspondence with the
one-transit walk mentioned above.

\section{Conclusions and perspectives}
In order to explain the origin of singularities of the thermodynamic 
potentials in some first-order
equilibrium phase transitions, Lee and Yang proposed to examine the
behaviour of the zeros of the partition function. Such an idea turned
out to have far reaching consequences. Indeed, subsequent
generalizations of this approach showed that from the analysis of
zeros of the partition function one can extract a lot information
about phase transitions in many different systems. When combined
with finite-size scaling arguments, this method may be used
to calculate the location of the transition point
and even critical exponents. An important point is the fact
that in this approach we directly refer to the properties of the
partition function, which is the most basic quantity of equilibrium
statistical mechanics. Thus, one can apply mathematically rigorous
techniques to obtain a number of interesting results concerning,
e.g., the very existence and the location of phase transitions in
various systems.

Recent large interest in the Yang-Lee approach is
related, however, to its further extension to nonequilibrium
phase transitions. Several examples show that accomplishing  such an
ambitious task may be feasible in some nonequilibrium systems. 
But  the entire approach is much more
problematic than in equilibrium situations. 
It is not entirely clear what is the analogue of the
partition function in this case. For the directed percolation,
coalescence of zeros was observed for the percolation probability --
that is actually the order parameter of the system. For
self-organized criticality this effect appears for some generating
functions. Perhaps the most promising are the results obtained for TASEP
models. In this case the normalization factor of a steady-state
probability distribution seems to play the role of the partition
function. In some cases one can even establish much closer
relations with equilibrium systems. But it is not known yet whether
such an approach can be applied to other classes of nonequilibrium
systems. For example, it would be interesting to test this method
on systems for which the transition rates are not constant, but
state-dependent, like, e.g., the  zero-range processes (ZRP).

It seems therefore  that, after more than fifty years of generalizations and
extensions, the full potential of the Yang-Lee approach is yet to be
uncovered. Important developments in its application to the study of nonequilibrium
phase transitions are in progress.

\section*{Acknowledgements}

We are grateful to Dr. Fran\c{c}ois Coppex for precious help in
preparing the manuscript and acknowledge partial support from the Swiss
National Science Foundation. A.L. acknowledges the support from
KBN through the research grant 1 P03B 014 27.

\end{document}